\documentclass{aa}
\usepackage{amssymb,amsfonts,amsmath,times,pict2e}
\usepackage{natbib}
\usepackage{graphicx}
\usepackage{enumerate}
\usepackage{subfigure}
\usepackage{bm}
\usepackage{longtable}
\usepackage{color}
\usepackage{ulem}
\usepackage{media9}
\definecolor{mygray}{gray}{0.6}
\definecolor{magenta}{rgb}{0.858, 0.188, 0.478}
\definecolor{brown}{rgb}{0.5, 0.15, 0.0}
\usepackage{xspace}

\newcommand{\fg}[1]{Fig.~\ref{fig:#1}}
\newcommand{\Fg}[1]{Figure~\ref{fig:#1}}
\newcommand{\eq}[1]{Eq.~(\ref{eq:#1})}
\newcommand{\Tb}[1]{Table~\ref{tab:#1}}

\newcommand{\se}[1]{Sect.~\ref{sec:#1}}

\newcommand{\eg}{e.g.,\xspace}
\newcommand{\ie}{i.e.,\xspace}

\newcommand{\sioo}{SiO$_2$\xspace}
\newcommand{\hho}{H$_2$O\xspace}
\begin{document}
\title{How planets grow by pebble accretion}
\subtitle{III. Emergence of an interior composition gradient}
\author{Chris W. Ormel\inst{1}, Allona Vazan\inst{2,3}, Marc G. Brouwers\inst{4}}
\institute{Department of Astronomy, Tsinghua University, Beijing 100084, China
\label{inst1}
\\
\email{[chrisormel@tsinghua.edu.cn]}
\and
Department of Natural Sciences, The Open University of Israel, Ra'anana 43537, Israel
\label{inst2}
\and
Astrophysics Research Center of the Open University (ARCO), The Open University of Israel, Ra'anana 43537, Israel
\label{inst2a}
\and
Institute of Astronomy, University of Cambridge, Madingley Road, Cambridge CB3 0HA
\label{inst3}
}    

\date{\today}

\abstract{During their formation, planets form large, hot atmospheres due to the ongoing accretion of solids. It has been customary to assume that all solids end up at the center constituting a "core" of refractory materials, whereas the envelope remains metal-free. Recent work, as well as observations by the JUNO mission, indicate however that the distinction may not be so clear cut. Indeed, small silicate, pebble-sized particles will sublimate in the atmosphere when they hit the sublimation temperature ($T\sim2\,000$ K). In this paper we extend previous analytical work to compute the properties of planets under such a pebble accretion scenario. We conduct 1D numerical calculations of the atmosphere of an accreting planet, solving the stellar structure equations, augmented by a non-ideal equation of state that describes a hydrogen/helium-silicate vapor mixture. 
Calculations terminate at the point where the total mass in metal equals that of the H/He gas, which we numerically confirm as the onset of runaway gas accretion.
When pebbles sublimate before reaching the core, insufficient (accretion) energy is available to mix dense, vapor-rich lower layers with the higher layers of lower metallicity.  A gradual structure in which $Z$ decreases with radius is therefore a natural outcome of planet formation by pebble accretion. 
We highlight, furthermore, that (small) pebbles can act as the dominant source of opacity, preventing rapid cooling and presenting a channel for (mini-)Neptunes to survive in gas-rich disks. Nevertheless, once pebble accretion subsides, the atmosphere rapidly clears followed by runaway gas accretion. We consider atmospheric recycling to be the more probable mechanisms that have stalled the growth of these planets' envelopes.
}

\maketitle

\section{Introduction}
The core accretion paradigm postulates that planet formation is a two-staged process: first a "solid" core forms by accreting refractory material, followed, when the core has become sufficiently massive, by the pouring in of metal-poor gas to form the envelope of a (giant) planet \citep{PollackEtal1996,HubickyjEtal2005,MordasiniEtal2012}. However, the assumption of a sharp division between metal-rich assembly for the core and metal-poor assembly for the gaseous envelope is problematic, as these two components form together. 
A planet first begins to bind a small proto-atmosphere when the planet's Bondi radius exceeds the core radius, corresponding to a mass of $\sim$$0.1\,M_\oplus$ dependent on disk location. The atmosphere grows into a larger envelope as the planet gains in mass. In the classical scenario, this envelope is presumed to remain the same composition as the nebular gas. 
This assumption is now increasingly being questioned, as the accreting solids may end up to enrich the atmosphere (to some extent) before they reach the core. \citet{Stevenson1982} already highlighted the importance of atmosphere enrichment. By increasing the molecular weight, the scaleheight of the atmosphere is decreased, allowing the envelope to contract and to become more massive.  More recently, \citet{HoriIkoma2011} and \citet{VenturiniEtal2015,VenturiniEtal2016} have argued that envelope enrichment by \hho would greatly reduce the critical core mass -- the mass where gas accretion accelerates.  

Since the amount of envelope enrichment is crucial to the thermodynamical evolution of the planet, the question arises what the fate of the solids is as they traverse through the envelope towards the core. In the classical model, any impactors that enter the envelope are assumed to remain intact until they reach the core, where they add their mass and liberate their gravitational energy. This requires that impactors are strong enough to overcome disruption due to the dynamical pressure they experience in their descent through the envelope \textit{and} large enough to avoid thermal ablation. In practice, this requires impactors to be rigid (rubble piles will disrupt) and to exceed $\sim$10\,km with this size threshold increasing with planet mass \citep{PodolakEtal1988,MordasiniEtal2015,PinhasEtal2016,BrouwersEtal2018,VallettaHelled2019,BrouwersOrmel2020}. The assumption that solids reach the core therefore becomes especially questionable when they are small

In the last decade, pebble accretion has emerged as a new contender to make large cores \citep{OrmelKlahr2010,LambrechtsJohansen2012,BitschEtal2015,JohansenLambrechts2017}. Pebbles, by virtue of their small size, settle gently towards the core. Therefore, it can be assumed that they are in thermal equilibrium with the gas -- that is, their sublimation is regulated by the saturation vapor pressure of the material. They will evaporate entirely when the ambient temperature exceeds the condensation temperature -- around $\sim$300\,K for \hho or $\sim$2\,000\,K for silicates.

In \citet[][henceforth Paper I]{BrouwersEtal2018} we numerically calculated the point where silicate impactors fully evaporate in the atmosphere and where direct core growth terminates. Importantly, we relaxed the assumption that enrichment is uniform in the envelope. Instead, the saturation vapor pressure limits the atmosphere intake of pollutants, which renders the enrichment strongly inhomogeneous. The outer layers are barely affected and remain dominated by hydrogen and helium, whereas the hot regions close to the core first become dominant in vapor. Of course, the nature of the enrichment depends on the materials from which the accreted solids are composed (rock vs volatile). In this work, we focus on silicate-rich impactors. With increasing mass, the envelope becomes hotter and is capable to absorb an increasingly amount of vapor, until solids no longer reach the core -- the end of the direct growth phase (Phase I).

With an analytical model, \citet[][henceforth Paper II]{BrouwersOrmel2020} described the subsequent evolution of such vapor-rich planets. In Phase II, growth is dominated by the envelope, which vaporizes and absorbs all incoming solid material. At the same time hydrogen/helium nebular gas is accreted and we gave an expression for the critical \textit{metal mass} -- the mass where the total amount of metals (core+vapor) equals the mass in hydrogen/helium. However, in case of (sub-)Neptunes and lower mass planets, the envelope never reached this point. Accretion of pebbles may shut off at some point, in which case the envelope evolves through Kelvin-Helmholtz contraction (Phase III). Finally, after the planet has emerged from the natal disk and has cooled for $\sim$Gyr (Phase IV), conditions may again become suitable for supersaturation and a renewed phase of core growth.

While the (semi-)analytical modelling provide a useful roadmap, they necessarily include assumptions to make them mathematically tractable. Specifically, we employed the ideal equation of state and (for Paper II) made the assumption that all vapor was confined in an inner homogeneous layer, which is convective. The caveat to the first assumption is that the rather compressible ideal gas results in very high vapor densities, while the assumption of a convective interior implies the presence of energy to mix the vapor with the hydrogen/helium gas. Numerical calculations are necessary to reexamine the conclusions reached in our previous works. 

To this end, we will employ a non-ideal, tabulated equation of state (EoS) both for the hydrogen/helium \citep{SaumonEtal1995} as well as for the \sioo vapor \citep{FaikEtal2018}. We calculate the EoS of the three component mixture self-consistently at each location in time following \citet{VazanEtal2013}. Our numerical model switches to a wet adiabat in regions where the vapor is saturated. 
In particular, we relax the assumption that the interior is entirely convective or conductive. Instead, we  condition convective transport and mixing with the appearance of positive luminosity, because mixing stratified \sioo vapor consumes energy.

We then apply this model to understand the origin of close-in super-Earth and mini-Neptune planets, which our galaxy harbors in great numbers \citep{FressinEtal2013,PetiguraEtal2013,ZhuEtal2018}, from a pebble accretion perspective. From arguments based on evaporation, the consensus is that these planets were born early, in the gas-rich phase, made out of terrestrial materials with a thick hydrogen/helium atmosphere \citep{WuLithwick2013,LopezFortney2014,FultonEtal2017,OwenWu2017}.  This could simply reflect a compact and dense progenitor disk \citep{ChiangLaughlin2013}, or indicated large-scale transport of either the planets themselves or of the planet building blocks \citep{Hansen2009}.  The case for pebble accretion is attractive, as the pebble flux can be maintained over a long time \citep{LambrechtsEtal2014,LambrechtsEtal2019} and, unlike with planet migration, a rocky composition can be expected as the \hho vapor will sublimate off pebbles when they cross the disk iceline.  Furthermore, pebbles may contribute to the atmospheric opacity, which determines the thermal evolution of the planet \citep{IkomaEtal2000,HoriIkoma2010,PisoYoudin2014}. We explore the possibility that pebbles in this way regulate the envelope growth.

A second application concerns the distribution of the vapor within the envelope.  Recently, new insights have emerged from Jupiter's gravitational moments measurement by the JUNO mission. These infer that, instead of the classical pure refractory core, the core is extended or ``dilute'' \citep{WahlEtal2017,DebrasChabrier2019}.  
The influence of the initial metallicity profile of the interior of (giant) planets on its long-term evolution has now been well studied \citep{VazanEtal2016,VazanEtal2018,LozovskyEtal2017,MuellerEtal2020}. Our previous works have already highlighted the potential for producing such dilute (vapor-rich) regions from formation.  It is therefore time to outline what different planet formation models predict for the internal structure of various types of planets -- giants as well as (mini-)Neptunes \citep[cf.][]{BodenheimerEtal2018,VallettaHelled2020,VenturiniHelled2020}.

The plan of the paper is as follows. In \se{model} we outline the model. Our model solves the stellar structure equations using a boundary element solver. In \se{results} we present our results for about one hundred simulations where we vary a great number of parameters. We present profiles of temperature, density, and metallicity, and compare some of our results to analytical predictions of Paper II. In \se{discuss} we discuss the potential for pebble accretion to explain the formation of (mini-)Neptune planets and dilute cores.  We list our conclusions in \se{conclude}.

\section{Model}
\label{sec:model}
Our model calculates the evolution of the core and envelope of an embedded, pebble-accreting planet. It includes elements from approaches applicable to planetesimal accretion \citep{Stevenson1982,PollackEtal1996,Rafikov2006,MordasiniEtal2012}, as well as elements from studies that consider pure Kelvin-Helmholtz contraction \citep{IkomaEtal2000,PisoYoudin2014,ColemanEtal2017,GuileraEtal2020}. The envelope is assumed to be in hydrostatic equilibrium on dynamical timescales and in pressure equilibrium with the surrounding disk gas. This validates a quasi-steady approach. At each time, or snapshot, a steady solution to the stellar structure equations is found.  

A key element of our approach is that we relax the standard assumption that accreting solids end up in the core. In Paper I we calculated the fraction of the incoming solids that end up in the envelope as silicate vapor -- a number steadily increasing with time until all the accreted pebble flux could be absorbed by the atmosphere.  Here, we extend these calculations to the undersaturated state, which requires a model for the treatment of the silicate vapor.

\subsection{Solid and vapor treatment}
We consider a growing protoplanet, embedded in a disk, accreting pebbles at a rate $\dot{M}_\mathrm{peb}$. These pebbles partially or entirely vaporize in the envelope of the planet. 
Formally, the evolution of the vapor distribution is described by a (time-dependent) transport equation, accounting for mixing, deposition (through ablation) and rainout \citep[\eg][]{OrmelMin2019}. 
Usually, transport is ignored, however, and the enrichment is for simplicity taken to be uniform. 
Nevertheless, in reality we can expect a stratified picture, reflecting the different condensation temperatures of volatile and refractory species \citep{IaroslavitzPodolak2007,LozovskyEtal2017}.  
In this work, we do account for the inhomogeneous nature of enrichment, but for simplicity consider a single refractory species (\sioo).
Let the density of the vaporized \sioo be denoted $\rho_Z$ and that of the non-condensible hydrogen/helium gas be $\rho_\mathrm{xy}$.  The metallicity of the gas is then $Z=\rho_Z/\rho = \rho_Z/(\rho_Z+\rho_\mathrm{xy})$.

In Paper I, we covered the early phases of sublimation, where the vapor follows the saturation pressure, which is a strong function of temperature. Consequently, the vapor density profile is non-homogeneous. In the analytical model of Paper II we added an inner layer, below the saturated (cooler) region, where the temperature is hot enough for the vapor to be below saturation. In this layer (region 2a in \fg{Zsketch}) the vapor resides at uniform concentration, as it was assumed that the layer is convective. With time the vapor concentration in this region $Z_\mathrm{cnv}$ decreases with the accretion of nebular gas. The assumption that silicate vapor and nebular gas mix, however, requires energy, which we found was not always available. This implies that the composition remains stratified to a certain degree.  In order to incorporate the limitations of compositional mixing, we add a third region (2b) where the composition is inhomogeneous in space but constant in time; the given mass shell is said to be "compositionally frozen" \textit{in time}. The structure of the envelope is sketched in \fg{Zsketch}. In summary, the regions are: 
\begin{enumerate}
    \item an outermost \textbf{saturated region}, where the composition follows the equilibrium metallicity
        \begin{equation}
            Z = Z_\mathrm{sat} = \frac{\rho_\mathrm{sat}}{\rho_\mathrm{sat} +\rho_\mathrm{xy}}
            \label{eq:Zeq}
        \end{equation}
    \item[2a.] a well-mixed \textbf{convective region} of uniform composition with $Z$ uniform and decreasing with time, $Z=Z_\mathrm{cnv}(t)$
    \item[2b.] a stratified, \textbf{frozen region} where the composition spatially varies, but where $Z$ of a particular mass shell does not change with time, $Z=Z(m)$.
\end{enumerate}
Our model design regarding composition can be regarded as an improvement over a pure global treatment, where $Z$ would just be constant throughout the envelope. In our model, the composition is, respectively, quantified by the value of $Z$ at previous times (region 2b), the constant $Z_\mathrm{cnv}$ (region 2a), and $Z_\mathrm{sat}$, which is a function of temperature.  
The compositional gradient attains their steepest values in region 1, just near the interface with region 2. Region 1 is by definition saturated. In saturated regions the \citet{Ledoux1947} convection criterion is not applicable, as the composition and temperature are no longer independent \citep[e.g.][]{Chambers2017}; the instability criterion is that of moist convection \citep{TremblinEtal2019}, which we include as a modified Schwarzschield criterion \se{wet-adiab}. Therefore, the model design choices avoid the explicit use of the Ledoux criterion.

The boundaries between these regions evolve with time and are set by the following conditions: $Z_\mathrm{cnv}=Z_\mathrm{sat}$ for the boundary between region 1 and 2a; and $L>0$ globally for the boundary between regions 2a and 2b. (Note again that these regime boundaries only appear in Phase II, see below). With time, both boundaries shift outward. The global condition for the luminosity requires an iterative approach -- we increase $r_\mathrm{iso}$ until $L$ becomes globally positive.

We next review how these regions develop chronologically and how this is implemented in our code. 
\begin{figure}[t]
    \includegraphics[width=\columnwidth]{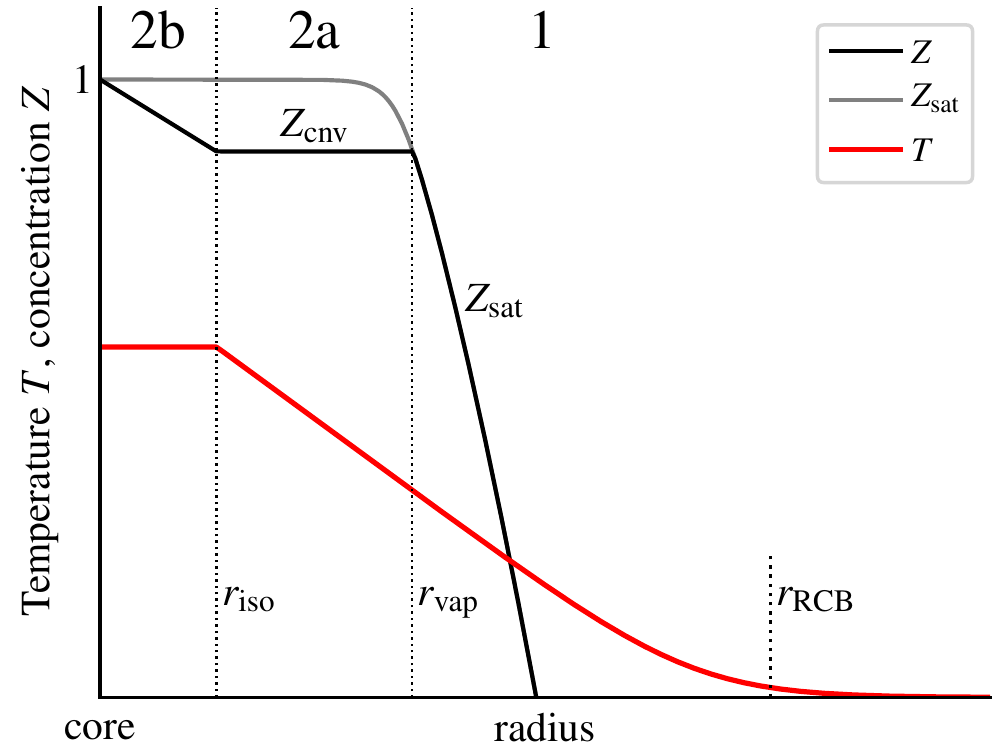}
    \caption{Adopted model for the metallicity $Z=\rho_Z/\rho$ of the envelope. In region 1 the envelope is always saturated and the metallicity $Z$ follows the saturation curve $Z_\mathrm{sat}$ (gray), which is set by the saturation vapor pressure. Initially, when the entire envelope is saturated, only this region is present. Upon reaching Phase II a new region (2) develops that is undersaturated ($Z<Z_\mathrm{sat})$. In region 2a convection ensures that the metallicity $Z_\mathrm{cnv}$ is uniform. Region 2b represents a zone where convection does not operate due to lack of energy sources. Here $Z$ is constant in time for the same mass shell. As more and more H/He gas is accreted, $Z_\mathrm{cnv}$ decreases and the region boundaries shift outwards.}
    \label{fig:Zsketch}
\end{figure}

\subsubsection{Phase I: Direct core growth}
In Phase I only region 1 exist. During this time the planet's envelope is sufficiently small and cold for refractories to survive thermal ablation and to make it to the core. However, the envelope absorbs as much vapor as it can absorb -- \ie region 1 is saturated -- and the amount of ablation increases with depth in the atmosphere, as the saturation vapor pressure is a strong function of temperature. 
This pressure can also be cast in terms of an equilibrium density $\rho_\mathrm{sat}=P_\mathrm{sat}/v_Z^2$ where $v_Z$ is the thermal velocity of \sioo. Hence, we assume in region 1 that $\rho_Z = \rho_\mathrm{sat}$ and define $Z_\mathrm{sat}$ accordingly. 

Initially, the amount of vapor absorbed by the envelope is small since $T$ is low; most of the pebbles end up in the core. When the temperatures approaches $\sim$2,000\,K sublimation becomes more significant and pebbles by virtue of their small sizes may entirely evaporate. Still, the fraction of vapor that the envelope can absorb is always limited to $\rho_\mathrm{sat}$. In Paper I the pebble's impact trajectory and vapor ablation rate was calculated explicitly. When found that $\rho_z>\rho_\mathrm{sat}$, the excess vapor was assumed to rainout to the core. In this work we omit an impact model and simplify matters by assuming that solids always settle slowly enough for the atmosphere to become saturated at any point. This assumption is natural to pebbles. The duration of Phase I lasts until the point where the envelope has become hot enough to become sub-critical ($\rho_\mathrm{sat}(T)>\rho_z$). All silicates material dissolves and core growth terminates.

\subsubsection{Phase II: Envelope growth}
The subsequent Phase II describes the phase where pebbles completely evaporate in the envelope. Core growth terminates and the envelope becomes undersaturated, meaning that $Z<Z_\mathrm{sat}$ in the hotter regions (2a and 2b), while in the cooler outer region 1 $Z=Z_\mathrm{sat}$ is maintained. The metallicity of the convective zone $Z_\mathrm{cnv}$, a solution unknown (see \se{bc-sol}), decreases with time. In Phase II this new parameter is substituted for the now known (fixed) core mass.

Phase II is also characterized by the development of the compositionally frozen region 2b. The boundary between these regions 2a and 2b, $r=r_\mathrm{iso}$, is another (unknown) parameter to the model. The condition we adopt is to demand that $L\ge0$ everywhere. If this is not satisfied, we let $r_\mathrm{iso}$ expand until $L$ becomes positive. In general, the region within $r_\mathrm{iso}$ releases energy due to $PdV$ compression, while mixing that takes place in region 2a consumes energy. The requirement that $L>0$ for mixing (or, rather, that $\nabla_\mathrm{rad} \gg 1$ wherever $L$ is positive in region 2a) may be replaced by a more stringent criterion amounting to, $L>L_\mathrm{min}$. But this would simply amount to shifting the 2a/2b boundary.

Our choice for this rigid distinction between regions 2a and 2b warrants explanation. The more desired choice is for a local approach, \eg\ to freeze the composition wherever the luminosity becomes too small for convective transport. There are reasons, mainly computationally in nature, however, forcing us to opt differently. First, the local approach may results in many mutually isolated convective zones, each with different $Z_\mathrm{cnv}$. More problematic, however, is that the numerical solver we have adopted simply cannot handle steep variations (let alone discontinuities) in the temperature gradient $\nabla$, which arise when the luminosity becomes negative -- the adiabatic index drops from a value consistent with convective transport (when $L>0$), $\nabla\approx$0.2--0.5 to $\ll$$-1$ when $L<0$. So, even though $L$ may change gradually, $\nabla$ will not. 
Therefore, we make the approximation that the structure is layered, characterized by one single "frozen" region. Actually, we find that within this region, the luminosity is generally positive, because the $PdV$ liberated heat from compression tends to outpace the increase in the internal energy.  Still, this luminosity is insufficient to mix the \textit{entire} regions 2a and 2b. Presumably, in reality multiple small convective layers develop, each with a different composition \citep{RosenblumEtal2011}. But our numerical tool is not suited to implement such a multi-layer model effort.

The extent of the compositionally frozen zone is such that it (collectively) generates positive luminosity needed to mix the material above. For simplicity the zone is modeled as isothermal. Keeping an adiabatic profile (apart from being inconsistent with the frozen assumption) requires large amounts of energy to support the high central temperatures. A negative $\nabla$ for the frozen zone, on the other hand, implies that there is a net energy transport into the core. While this may partially develop because our core is of lower temperature, the ability of the core to transport energy in the direction of gravity (\ie\ no convective transport!) is questionable. Taking the region as a whole, the isothermal assumption can be justified as there are no obvious sources of energy (pebbles being deposited above this layer). Our situation therefore resembles the case in stellar evolution, where an isothermal stellar core develops after hydrogen burning moves away from the star's core due to it being exhausted of hydrogen \citep{KippenhahnWeigert1990}. Obviously, the assumption of $\nabla=0$ across the entire region is crude and a local treatment would be preferable. But in an averaged sense, it captures the spirit that a lack of energy sources causes $\nabla$ to decrease.

\subsubsection{Phase III: Embedded cooling}
\label{sec:phase3}
In Phase III pebble accretion terminates and gives way to pure Kelvin-Helmholtz contraction of the envelope. This phase is not associated with the introduction of a new region. We include Phase 3 simply by tapering off the pebble accretion rate: 
\begin{equation}
    \dot{M}_\mathrm{peb} = \dot{M}_\mathrm{peb,0} \exp \left[ -A \left( \frac{t}{t_\sigma}\right)^n \right]
\end{equation}
where $t_\sigma=M_{Z,\mathrm{final}}/\dot{M}_\mathrm{peb,0}$ is the timescale over which the pebble accretion rate subsides, $M_{Z,\mathrm{final}}$ the final metal mass of the planet, $A=(\Gamma((n+1)/n))^n$ a constant to make $\int \dot{M}_\mathrm{peb,0} dt = M_{Z,\mathrm{final}}$, and $\Gamma(x)$ the Gamma function. 

In most of our runs, we for simplicity take an infinite and constant supply of pebbles $M_{Z,\mathrm{final}} = t_\sigma=\infty$.  But in runs where we do explore Phase III we choose $M_\mathrm{Z,final}=4$, 6, and $8\,M_\oplus$ as representative for a mini-Neptune. We further choose $n=4$ ($A=0.675$) such that after $t=t_\sigma$ accretion of pebbles very rapidly subsides to zero. 

\subsubsection{Phase IV: Isolated cooling}
We have defined Phase IV as the phase where the protoplanetary disk dissipates, the outer layers of the atmosphere may evaporate, and where progressive cooling may cause the vapor to rainout. This phase is not considered in the present work, but is crucial to assess the viability of core-powered mass loss (\citealt{OwenWu2016,GuptaSchlichting2019}; Paper II). As it covers the $\sim$Gyr evolution, it also connects directly to the observed properties of exoplanets.

\subsection{Envelope structure equations}
We obtain the temperature ($T$) and pressure ($P$) profiles of the envelope by solving the stellar structure equations:
\begin{subequations}
    \label{eq:ss-all}
\begin{equation}
    \frac{\partial P}{\partial m} = - \frac{Gm}{4\pi r^4}
\end{equation}
\begin{equation}
    \frac{\partial T}{\partial m} = - \frac{Gm}{4\pi r^4} \frac{T}{P} \nabla
\end{equation}
\begin{equation}
    \frac{\partial r}{\partial m} = \frac{1}{4\pi r^2 \rho}
\end{equation}
\begin{equation}
    \label{eq:Lcnsv}
    \frac{\partial L}{\partial m} = -\frac{\partial u_\mathrm{int}}{\partial t} -P\frac{\partial V}{\partial t}
\end{equation}
\end{subequations}
where $\nabla = \min(\nabla_\mathrm{rad}, \nabla_\mathrm{ad})$ (Schwarzschield criterion) with $\nabla_\mathrm{rad}$ the radiative gradient
\begin{equation}
    \nabla_\mathrm{rad} = \frac{3 \kappa L P}{64\pi \sigma_\mathrm{sb} Gm T^4}
    \label{eq:nabla-rad}
\end{equation}
$\kappa$ the opacity, $L$ the luminosity, $\sigma_\mathrm{sb}$ Stefan-Boltzmann's constant, and $\nabla_\mathrm{ad}$ the (wet) adiabatic gradient, see \se{wet-adiab}. Following Paper I we use a simple power-law fit for the opacity, which we adopt from \citet{LeeChiang2015}: 
\begin{subequations}
    \label{eq:kappa-molecular}
    \begin{equation}
        \kappa_\mathrm{mol} = 7\times10^{-3}\,\mathrm{cm^2\,g^{-1}} \left(\frac{T}{10^3\,\mathrm{K}}\right)^{0.9} \left(\frac{\rho}{10^{-5}\,\mathrm{g\,cm^{-3}}}\right)^{0.3}
    \end{equation}
valid for our calculations at 0.2 au, and
    \begin{equation}
        \kappa_\mathrm{mol} = 1\times10^{-5}\,\mathrm{cm^2\,g^{-1}} \left(\frac{T}{10^2\,\mathrm{K}}\right)^{2.2} \left(\frac{\rho}{10^{-6}\,\mathrm{g\,cm^{-3}}}\right)^{0.6}
    \end{equation}
\end{subequations}
valid for 1 and 5 au. Expressions are fits to tables by \citet{FergusonEtal2005} and \citet{FreedmanEtal2014} and are evaluated at solar metallicity ($Z=0.02$). Sublimation will raise $Z$ but at this point envelopes have already become convective. 
As found by previous works grains are not expected to survive long in envelopes \citep{MovshovitzEtal2010,Mordasini2014,Ormel2014}. Instead, we account for an opacity from pebbles:
\begin{equation}
    \label{eq:kappa-peb}
    \kappa_\mathrm{peb} =
    \frac{3 Q_\mathrm{eff}}{4\rho_{\bullet\mathrm{peb}}s_\mathrm{peb}} \times \frac{\dot{M}_\mathrm{peb}}{4\pi r^2 v_\mathrm{sed} \rho}
\end{equation}
where $Q_\mathrm{eff}$ is the efficiency coefficient (which evaluates to 2), $s_\mathrm{peb}$ the radius of the pebbles, and $v_\mathrm{sed}$ is the sedimentation velocity. By assuming that gas drag and gravity balance, the sedimentation velocity of the pebble is $v_\mathrm{sed} = Gm t_\mathrm{stop}/r^2$.
In \eq{kappa-peb} the first term is the opacity per unit mass dust, the second term the dust-to-gas ratio, and the stopping time $t_\mathrm{stop}$ reads
\begin{equation}
    t_\mathrm{stop} = \frac{\rho_\bullet s_\mathrm{peb}}{v_\mathrm{th}\rho} \times \max \left[ 1, \frac{4s_\mathrm{peb}}{9l_\mathrm{mfp}} \right]
    \label{eq:tstop}
\end{equation}
where we account for the Epstein and Stokes drag law. Note that in \eq{kappa-peb} it is assumed that pebbles do not coagulate.

In addition to these equations, we integrate the metallicity $Z$ in order to obtain the total amount of solids in the envelope
\begin{equation}
    M_{z,\mathrm{env}}
    = \int_{M_\mathrm{core}}^{M_\mathrm{tot}} Z(m) \mathrm{d}m.
    \label{eq:Mz-eq}
\end{equation}
The total metal contents of the planets then changes as function of time
\begin{equation}
    \label{eq:mass-consv}
    M_\mathrm{core}+M_\mathrm{z,env} = \int \dot{M}_\mathrm{peb}(t) \mathrm{d}t
\end{equation}
which provides another boundary constraint.

In the above expressions, the gas density $\rho$, the internal energy $u_\mathrm{int}$ and the adiabatic index are obtained from an equation of state (\se{EoS}).

\begin{figure*}[t]
    \sidecaption
    \includegraphics[width=0.7\textwidth]{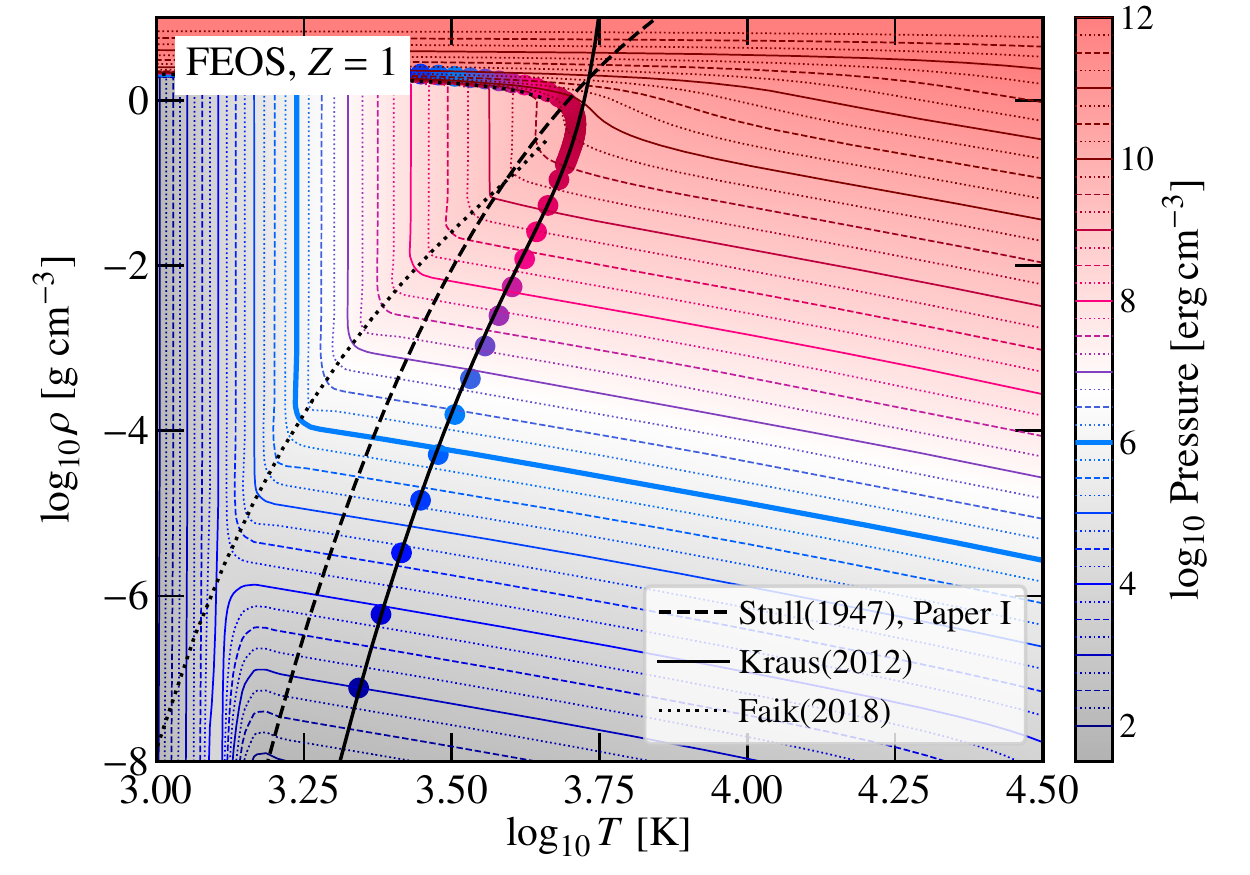}
    \caption{Frankfurt equation of state (FEOS). Shown are pressure contours as function of density and temperature for a fully \sioo composition ($Z=1$). Regions where the iso-pressure contour are vertical represent phase transitions. The liquid-vapor curve by \citet{Stull1947} (used in Paper I) is given by the dashed curve, while the colored dots indicate the experimentally inferred curve by \citet{KrausEtal2012} (the color indicates pressure). The fit to the \citet{KrausEtal2012} data (solid black curve) is used in this work as the liquid-vapor curve.}
    \label{fig:EoS-FEOS}
\end{figure*}
\subsection{Luminosity and core treatment}
\label{sec:luminosity-treatment}
We consider the energy conservation equation \eq{Lcnsv} in integral form:
\begin{equation}
    L(m) = -\frac{dE}{dt} -P\frac{dV}{dt} +L_s,
    \label{eq:Lint}
\end{equation}
where $L_s$ are surface contributions \citep{PisoYoudin2014}. In preceding works \citep{MordasiniEtal2012,PisoYoudin2014,VenturiniEtal2016} energy conservation was applied at a specific radius (e.g.\ the radiative-convective boundary) and the corresponding $L$ was assumed to be constant within this radius. In contrast, in this work \eq{Lint} is used to find the luminosity at every every mass shell, \ie\ $L(m)$ becomes a local quantity. The energy $E$ is then the total energy within shell $m$
\begin{equation}
    E = \Phi(m) +U(m)
      = E_\mathrm{core} +\int_{M_\mathrm{core}}^m \left( u_\mathrm{int} -\frac{Gm'}{r'}  \right) dm'
\end{equation}
    where $\Phi(m)$ is the potential energy at mass level $m$ and $U(m)$ is the total internal energy interior to $m$ and $E_\mathrm{core}$ the energy of the core. The internal energy includes the latent heat of vaporization. The core's potential energy is taken equal to that of a uniform sphere ($\Phi_\mathrm{core}= -3GM_\mathrm{core}^2/5R_\mathrm{core}$), while the core's thermal energy ($U_\mathrm{core}$) follows from integration over the temperature structure \textit{within} the core; that is,
\begin{equation}
    \label{eq:Ecore}
    E_\mathrm{core} = \Phi_\mathrm{core} +\int_0^{M_\mathrm{core}} c_V T(m) dm,
\end{equation}
where the heat capacity is taken to be $c_V=10^7\,\mathrm{erg\,K^{-1}\,g^{-1}}$ \citep{GuillotEtal1995,D'AngeloPodolak2015}. It is assumed that no thermal transport takes place within the core. Hence, $T(m)$ in \eq{Ecore} corresponds to the temperature at the bottom of the envelope at the time when pebbles became incorporated into the core. Indeed, pebbles, unlike planetesimals, accommodate to the ambient temperature (Paper I). The core is for simplicity treated as thermally isolated, that is, the further thermal evolution of the envelope will not affect it or vice-versa. We also do not use an equation of state for the core: it is instead characterized by a fixed density (\Tb{pars}).

\begin{figure*}[t]
    \centering
    \includegraphics[width=0.45\textwidth]{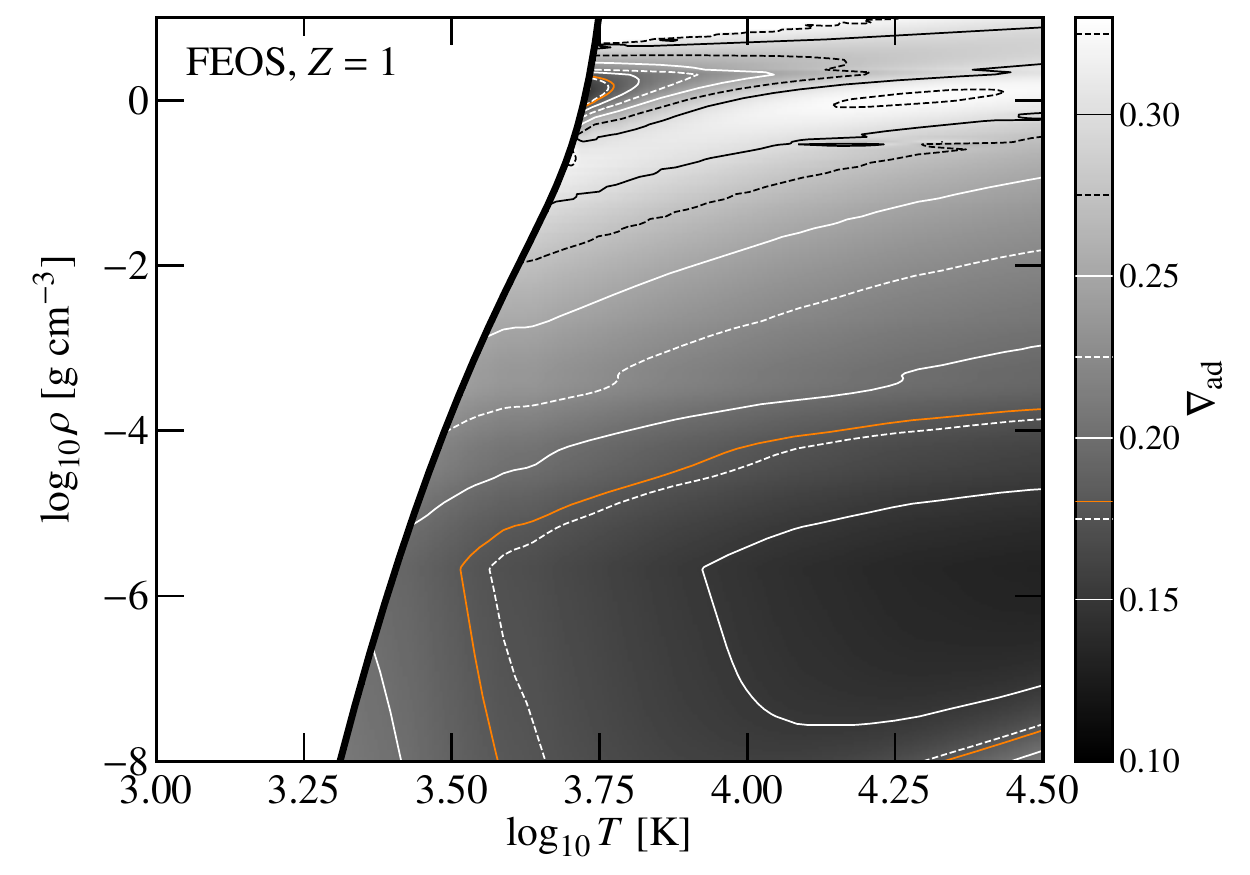}
    \includegraphics[width=0.45\textwidth]{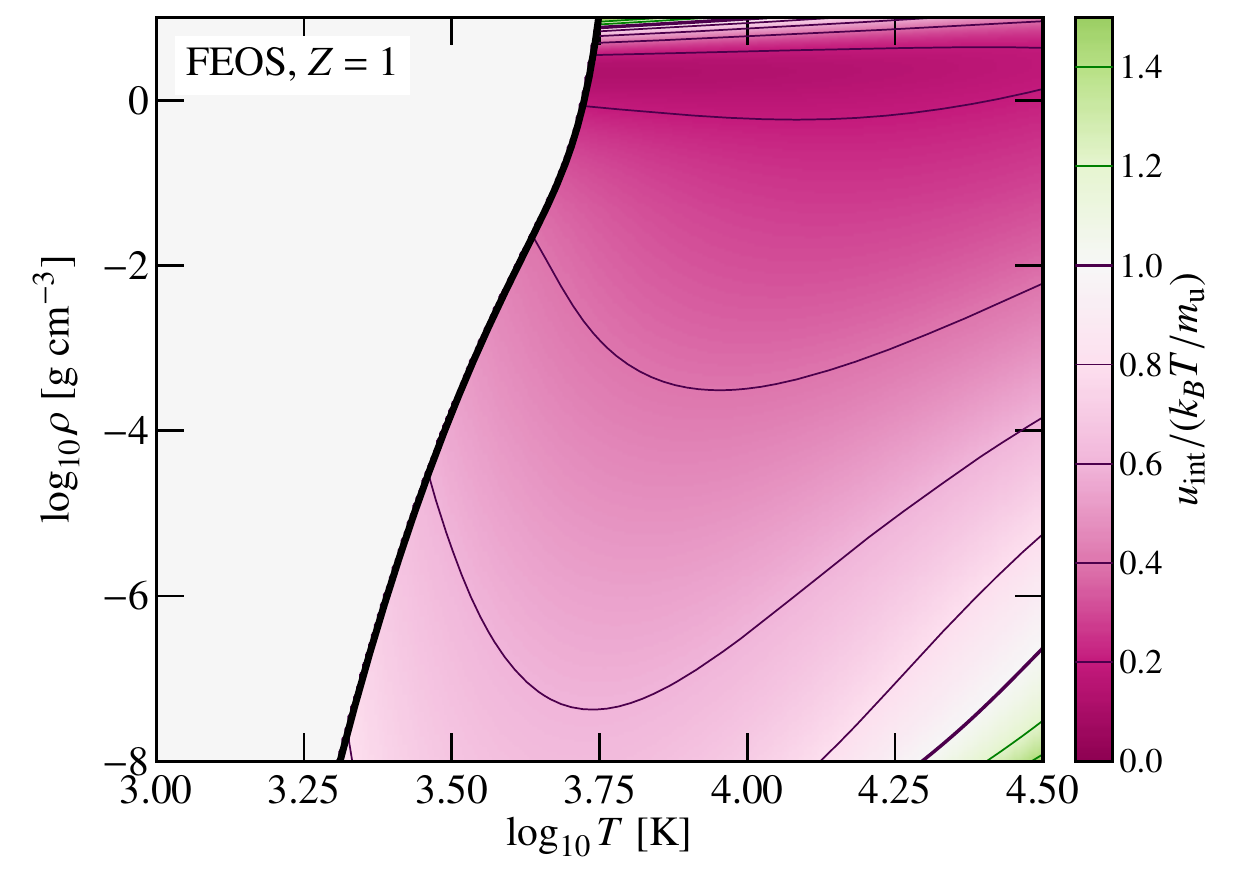}
    \caption{Frankfurt equation of state (FEOS) -- Adiabatic gradient and internal energy. For clarity, only data exterior to the Kraus liquid-vapor curve is shown. The orange curve in (a) denotes the value of $\nabla_\mathrm{ad}$ in the ideal limit. For the internal energy all values lie above the ideal limit of $u_\mathrm{int}=0.075 k_B T/m_\mathrm{u}$.}
    \label{fig:EoS-FEOS-other}
\end{figure*}

\subsection{Equation of state (EoS)}
\label{sec:EoS}
We assume that hydrogen and helium are present at a mass ratio of $X/Y=0.7/0.3=2.33$.  In this work, we consider both the ideal EoS and a tabulated, non-ideal EoS.

\subsubsection{Ideal EoS}
\label{sec:EoS-ideal}
In the ideal EoS the mean molecular weight $\mu$ of the gas mixture is
\begin{equation}
    \frac{1}{\mu} 
    = \frac{X}{2} +\frac{Y}{4} +\frac{Z}{\mu_Z}
    = \frac{(1-Z)}{2.35} +\frac{Z}{\mu_Z}
    \label{eq:mu-ideal}
\end{equation}
For \sioo $\mu_Z=60$. Dissociation and ionization of \sioo would reduce $\mu$, but only in regions where $Z\approx1$. 

Likewise the heat capacity ratio of the gas is obtained by the weighted ratio of the individual heat capacities of hydrogen (H$_2$), helium, and $Z$. We consider three translational degrees and two rotational degrees of freedom for the linear molecules $H_2$ and \sioo. Because \sioo becomes abundant at high temperatures, we also include its vibrational degrees (four). The total degrees of freedom for hydrogen, helium, and \sioo are 5, 3, and 9, respectively. The weighted heat capacity is then
\begin{equation}
    \gamma_\mathrm{ad} 
    = \frac{\sum_i C_{P,i}}{\sum_i C_{V,i}}
    = \frac{7X/2 +5Y/2 +11Z/2}{5X/2 +3Y/2 +9Z/2}
    = \frac{3.2 +2.3Z}{2.2 +2.3Z}
    \label{eq:gamma-ad}
\end{equation}
from which $\nabla_\mathrm{ad} = (\gamma_\mathrm{ad}-1)/\gamma_\mathrm{ad}$ follows.

The density and internal energy are:
\begin{equation}
    \rho = \frac{P \mu m_\mathrm{u}}{k_B T}; \qquad
    u_\mathrm{int} = \frac{1}{\gamma-1} \frac{k_B T}{\mu m_\mathrm{u}} +Ze_\mathrm{vap}
\end{equation}
where $e_\mathrm{vap}$ is the latent heat of evaporation -- the energy required to transform the solid pebbles to the gas phase. Hence, our $\gamma_\mathrm{ad}$ ranges from 1.45 in the (outer) H/He-dominated regions to $\gamma_\mathrm{ad}=1.22$ in region where metals dominate.

This ideal EoS is analytical and simplistic. Non-ideal effects as ionization and dissociation can in principle be included for hydrogen and helium \citep{IkomaEtal2000,LeeEtal2014}, while still keeping an analytical approach. We do not consider these refinements, in order to connect to our previous works of Paper I and II and to keep a clear separation to the non-ideal, tabulated EoS.

\subsubsection{Non-ideal EoS}
\label{sec:non-ideal}
We follow the method outlined in \citet{VazanEtal2013} to obtain the EoS of a mixture of hydrogen, helium, and \sioo. For the H/He mixture the SCVH EoS \citep{SaumonEtal1995}\footnote{Recently, the SCVH EoS was updated \citep{ChabrierEtal2019}, but the differences between the SCvH and the new EoS are minor in the parameter space of interests for this work.} is used at fixed helium-to-hydrogen mass ratio of $Y/X=0.3/0.7$. For \sioo we use the publicly-available Frankfurt equation of state (FEOS; \citealt{FaikEtal2018}), which uses a quotidian model \citep{MoreEtal1988} with a soft-sphere correction \citep{YoungCorey1995} to better represent the behavior near the critical point. We remark that the SCVH EoS includes H-dissociation and ionization, whereas the \sioo EoS only includes ionization. More generally, chemistry or interactions among materials in the mixture is not accounted for. The mixing with H/He follows the additive volume law \citep{VazanEtal2013}. The equation of state provides us with pressure and internal energy as function of density, temperature and metallicity $Z$.

Results of the FEOS non-ideal EoS ($Z=1$) are shown in \fg{EoS-FEOS}, where pressure contours are plotted as functions of temperature and density. Vertical lines indicate a phase transition between liquid and vapor: a small change in pressure is sufficient to change the density dramatically. However, beyond the critical temperature ($T_c = 5100\,\mathrm{K}$) a phase transition no longer takes place; the fluid becomes supercritical. Beyond the critical point, at higher densities the spacing between the iso-pressure curves decrease considerably: a supercritical fluid is hard to compress. This behavior is markedly different from the ideal case, that is, $P/P_\mathrm{ideal}\gg 1$.

In \fg{EoS-FEOS} the liquid-vapor curve (LV-curve) consistent with FEOS (dotted line) differs from the expression corresponding to \citet{Stull1947} (dashed curve), which we used in Paper I.  In addition, we overplot shock-and-release experimental data for the liquid-vapor transition \citep{KrausEtal2012}. It can be seen these data match FEOS quite well (\textit{i.e.}, the color of the dot, indicating the pressure from the experiments, matches the color of the iso-contour from the quotidian EoS) for the lower branch. The mismatch with FEOS regarding the upper branch is a simple consequence of the very different LV-curves. Again, we fit the LV-curve consistent with the lower branch of the \citet{KrausEtal2012}:
\begin{equation}
    \rho_\mathrm{sat,Kraus}
    = \exp\left[ a -\frac{b}{T} +\left( \frac{T}{T_c} \right)^c \right]
    \label{eq:rhosat-Kraus}
\end{equation}
(cgs units) where $a = 9.0945$, $b = 5.630\times10^4\,\mathrm{K}$, $c=13.26$ and $T_c=5130\,\mathrm{K}$ (sold curve). We fit the saturation vapor pressure with a similar expression.  We consider the \texttt{Kraus} liquid-vapor curve, parameterized by \eq{rhosat-Kraus} to be the most realistic. The \texttt{Stull} curve is also used, in order to facilitate comparison to Paper I.\footnote{For low temperatures, both the Stull and Kraus liquid-vapor curves lie to the right of the FEOS curve. With these choices we also avoid the ``problematic'' region around $\log_{10}T=3.1$ at low density, which is a numerical artefact that resulted from matching the EoS to the ideal gas law.}  Material in excess of the LV-curve is added to the core, which density is assumed to be fixed.

In \fg{EoS-FEOS-other} we also present the adiabatic gradient $\nabla_\mathrm{ad}$ and internal energy $u_\mathrm{int}$ normalized to $k_B T/m_\mathrm{u}$ for a vapor-only $Z=1$ fluid. In this way we contrast the thermodynamical properties of the non-ideal vapor with the ideal vapor gas, for which these values are about 0.18 ($\nabla_\mathrm{ad}$) and 0.075 ($u_\mathrm{int}$; not accounting for the energy of vaporization). As can be seen from \fg{EoS-FEOS-other} the adiabatic gradient is generally higher than ideal along the liquid-vapor curve and in the top part of this panel (these are the regions sampled by the simulations). The higher adiabatic gradient (lower compressibility)  at higher pressures reflects the fact that the molecular size becomes significant and that inter-molecular interactions become important. Higher temperatures imply more vibrational modes, and hence $\nabla_\mathrm{ad}$ tends to decrease in this direction. At low densities, ionization increases the internal energies of the gas.

\begin{table*}[tb]
\centering
\caption{\label{tab:pars}Model constants and boundary conditions.}
\small
\begin{tabular}{llrll}
    \hline
    \hline
    Parameter                   & Symbol    & Value & \\
    \hline
    Core density                & $\rho_\mathrm{core}$  & $\mathrm{3.2}$ & $\mathrm{g\,cm^{-3}}$ \\
    Core heat capacity          & $c_V$     & $10^7$ & $\mathrm{erg\,K^{-1}\,g^{-1}}$ \\
    Disk temperature$^a$        & $T_\mathrm{disk}$ &       [750, 350, 150] & K \\
    Disk density$^a$            & $\rho_\mathrm{disk}$ &    [$10^{-7}$, $2\times10^{-9}$, $5\times10^{-11}$] & $\mathrm{g\,cm^{-3}}$ \\
    Latent heat of vaporization$^c$ & $e_\mathrm{evap}$         & $1.52\times10^{11}$ &$ \mathrm{erg\,g^{-1}}$ \\
    Pebble accretion rate       & $\dot{M}_\mathrm{peb}$    & $[10^{-4}$, $10^{-5}$, $10^{-6}$, $10^{-7}$] & $\mathrm{M_\oplus\,yr^{-1}}$ \\
    Pebble internal density     & $\rho_\bullet$  & $\mathrm{2.0}$ & $\mathrm{g\,cm^{-3}}$ \\
    Pebble radius$^b$           & $s_\mathrm{peb}$          & [None, 0.01, 0.1, 1] & cm \\
    \hline
    \hline
\end{tabular}
    \tablefoot{
    \tablefoottext{a}{Given values indicative of distances 0.2, 1.0, and 5.0 from a 1 $M_\odot$ star.}
    \tablefoottext{b}{Radius only affects the opacity calculation.}
    \tablefoottext{c}{\citet{Melosh2007}}
}
\end{table*}

\begin{figure}[t]
    \includegraphics[width=\columnwidth]{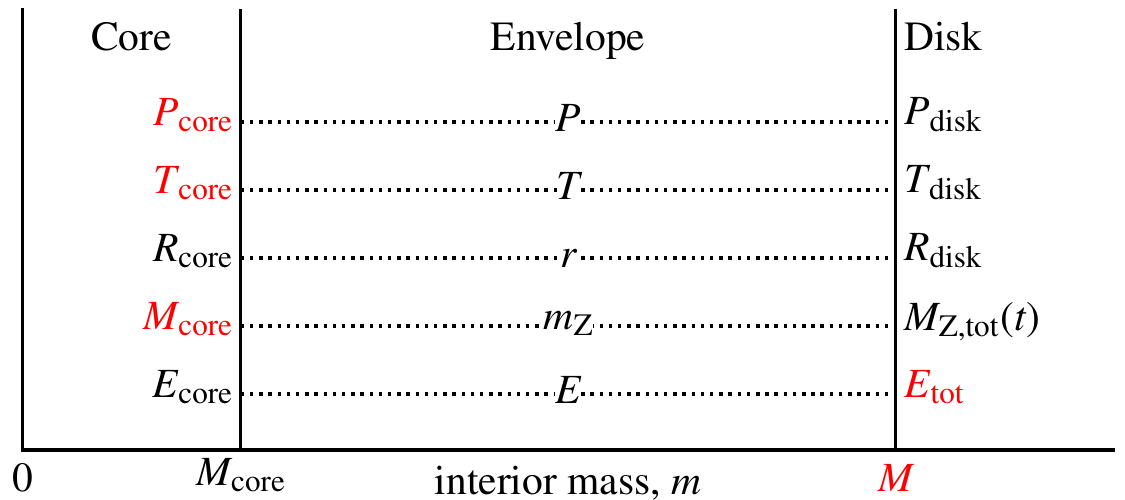}
    \caption{Schematic of the numerical problem. The (independent) grid parameter $m$ ranges from the core mass $M_\mathrm{core}$ to the (unknown) total mass $M$. The five structure equations for $P$, $T$, $r$, $M_z$, and $E$ are solved by imposing ten boundary conditions. The radii $R_\mathrm{core}$ and $R_\mathrm{out}$ follow from $M_\mathrm{core}$ and $M$, respectively, and $M_\mathrm{Z,tot}$ is incremented according to the timestep $\Delta t$, $\Delta M_\mathrm{Z,tot} = \dot{M}_\mathrm{peb}\Delta t$. The total energy $E$ is needed to provide the luminosity $L$ at any point (\eq{Lint}) . In Phase II $M_\mathrm{core}$ becomes a known parameter, while the metallicity threshold parameter $Z_\mathrm{max}$ is added as an unknown. }
    \label{fig:ssketch}
\end{figure}

\subsubsection{Wet adiabat treatment}
\label{sec:wet-adiab}
In the regions where the atmosphere becomes saturated, the density and pressure follows the liquid-vapor curve. This "wet" adiabatic gradient is therefore
\begin{equation}
    \nabla_\mathrm{wet}^{-1} = \frac{d\log P_\mathrm{sat}}{d\log T} = \frac{b}{T} +c \left(\frac{T}{T_c}\right)^c
\end{equation}
where the saturation vapor pressure $P_\mathrm{sat}$ is fit by the same expression as \eq{rhosat-Kraus} with $a=32.599$, $b=5.701\times10^4$ and $c=3.395$. Between 3,000 and 5,000 K, $0.05 < \nabla_\mathrm{wet} < 0.07$, much lower than $\nabla_\mathrm{ad}$ in the ideal EoS (0.18) and the non-ideal EoS. The lower adiabatic gradient arises because energy goes into sublimating the condensates at the expense of internal energy of the vapor.

In general the value of $\nabla_\mathrm{ad}$ will lie between the adiabatic index of a non-saturated mixture, provided by the EoS, and $\nabla_\mathrm{wet}$. To our knowledge, analytical expressions for such a mixed $\nabla_\mathrm{ad}$ is only available in the limit where the non-condensible gas is ideal \citep{Kasting1988,Chambers2017}. We therefore adopt a simple weighting:
\begin{equation}
    \label{eq:eos-mixed}
    \nabla_\mathrm{ad}^{-1} = w\nabla_\mathrm{wet}^{-1} +(1-w)\nabla_\mathrm{ad,eos}^{-1} 
\end{equation}
where the weighting factor is given by $w = (P_z/P_\mathrm{sat}) \times (P_z/P)$. The form of \eq{eos-mixed} and the choice of $w$ loosely follow \citet{Kasting1988} and \citet{Chambers2017}.\footnote{Note that their formula only applies to saturated regions (our region 1), where $P_z=P_\mathrm{sat}$ and $w=P_z/P$. We have added the factor $P_z/P_\mathrm{sat}$ to our weighting function to render it general and continuous. Therefore, as the vapor pressure steeply rises with temperature, the unsaturated region 2 will have $w=0$ and follow the dry $\nabla_\mathrm{ad}$.} Only when the vapor is saturated (region 1: $P_z=P_\mathrm{sat}$) \textit{and} when the vapor actually dominates the mixture ($P_z=P$) does the adiabatic gradient follows the wet adiabat. Conversely, when the vapor is saturated but is insignificant in terms of mass (at low temperature in region 1, $P_z \ll P$) or when the vapor is no longer saturated (high temperatures of region 2, $P_\mathrm{sat} \gg P_z$) we get the standard results from the mixed equation of state of two gases. We find that the reduction of $\nabla_\mathrm{ad}$ in Phase I is important as it ameliorates the positive feedback effect that drive the cusps of high temperature, density, and metallicity (Paper I and below). The wet adiabat treatment is applied in both the ideal and non-ideal simulations. 

\subsection{Boundary equations and solution technique}
\label{sec:bc-sol}
We choose time $t$ as the evolutionary parameter and increment $t$ in such a way to keep the relative increase in total mass limited to a few percent. The total mass $M_\mathrm{tot}$ is therefore an unknown parameter. The grid parameter is the internal mass $m$, which ranges from the core mass $M_\mathrm{core}$ at the bottom of the envelope to $M_\mathrm{tot}$ at the outer boundary.

In total the system consist of five differential equations, for pressure, temperature, radius, vapor mass, and energy as function of mass $m$. In addition we have the unknown parameters $M$ and $M_\mathrm{core}$ (Phase I) or $M_\mathrm{tot}$ and $Z_\mathrm{cnv}$ (Phase II). In order to solve the system boundary conditions (or relations among them) must therefore be specified.

This is shown schematically in \fg{ssketch}.  At the inner boundary of the envelope we take $r=R(M_\mathrm{core})$, where we assume that the core density is fixed at $\rho_\mathrm{core}$. Furthermore, $E=E_\mathrm{core}$, and $m_Z=M_\mathrm{core}$ are known. At the outer boundary we assume pressure and thermal equilibrium with the background disk, characterized by pressure $P_\mathrm{disk}$ and temperature $T=T_\mathrm{disk}$. The classical choice is to place the interface of disk and envelope at the minimum of the Bondi radius and the Hill radius:
\begin{equation}
    R_\mathrm{disk} = \min(R_\mathrm{Bondi}, R_\mathrm{Hill}).
    \label{eq:Rdisk}
\end{equation}
where $R_\mathrm{Bondi} = GM_p/\gamma k_B T$ and $R_\mathrm{Hill} = r (M_p/M_\star)^{1/3}$.  This closes the system of equations.

However, hydrodynamical studies indicate that the atmosphere may not be bound within the Bondi (or Hill) \citep{OrmelEtal2015i,FungEtal2015}. Instead, they hint at a complex flow pattern, where especially the outer atmosphere is continuously replenished with disk material, rendering cooling inefficient. The degree of this atmosphere-disk recycling is, unfortunately, somewhat complex, depending greatly on the ability of the gas to cool \citep{CimermanEtal2017,KurokawaTanigawa2018}, but the overall trend is that the outer regions are rapidly replenished, while the (very) inner regions are more prone to be bound. To encapsulate this effect in 1D evolutionary calculations, we may simply (and crudely) shift the outer radius inwards \citep{Ali-DibEtal2020}. Beyond this radius $R_\mathrm{adv}$ the gas may be modeled as adiabatic if cooling is insignificant on an advective crossing timescale: entropy is advected (and conserved). Assuming an ideal EoS for the gas, the conditions at $R_\mathrm{adv}$ are therefore:
\begin{equation}
    T_\mathrm{adv} = T_\mathrm{disk} \left( 1 +(\gamma-1) \left( \frac{R_B}{R_\mathrm{adv}} -\frac{R_B}{R_\mathrm{disk}} \right) \right)
    \label{eq:Tadv}
\end{equation}
and $P_\mathrm{out} = (T_\mathrm{adv}/T_\mathrm{disk})^{\gamma/(\gamma-1)}$ \citep{Ali-DibEtal2020}. In this work, we for simplicity fix $R_\mathrm{adv}=0.5R_\mathrm{disk}$ in some simulations to explore how entropy advection would affect the results.

\begin{figure*}
     \includegraphics[width=0.333\textwidth]{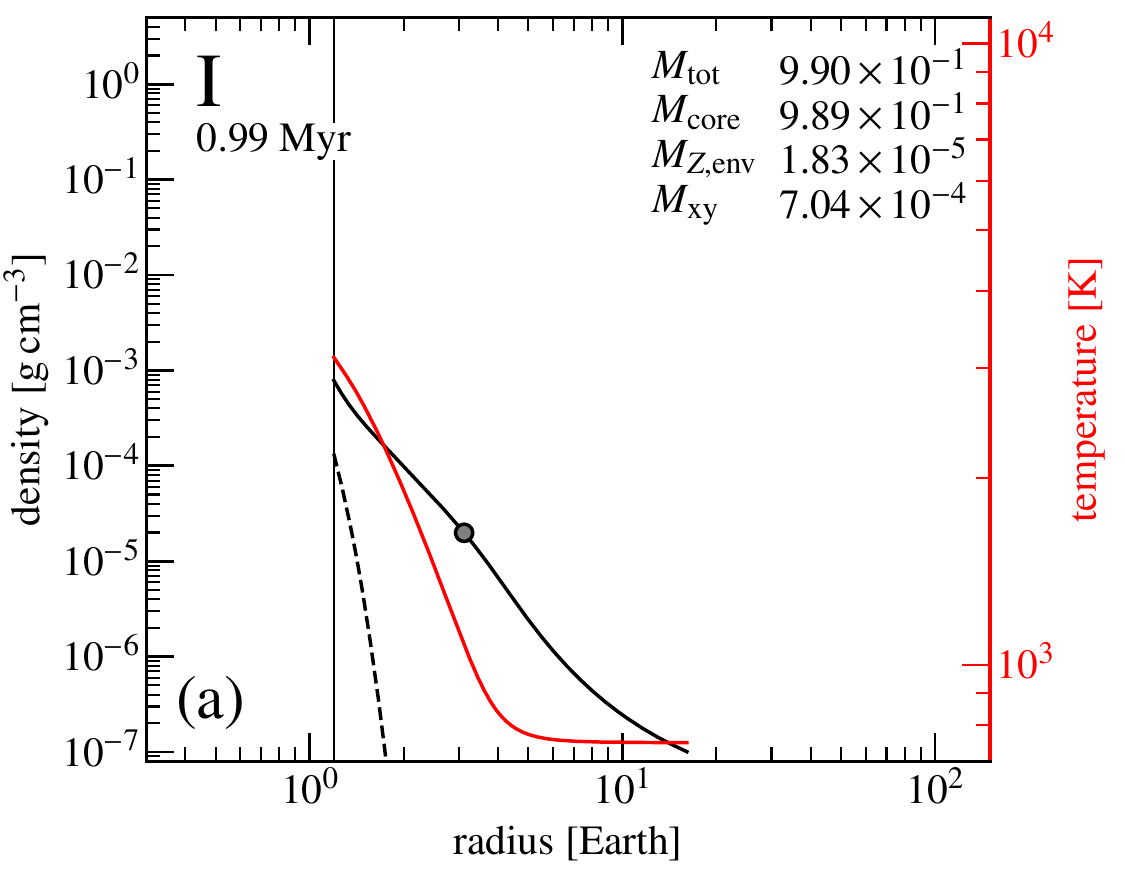}
     \includegraphics[width=0.333\textwidth]{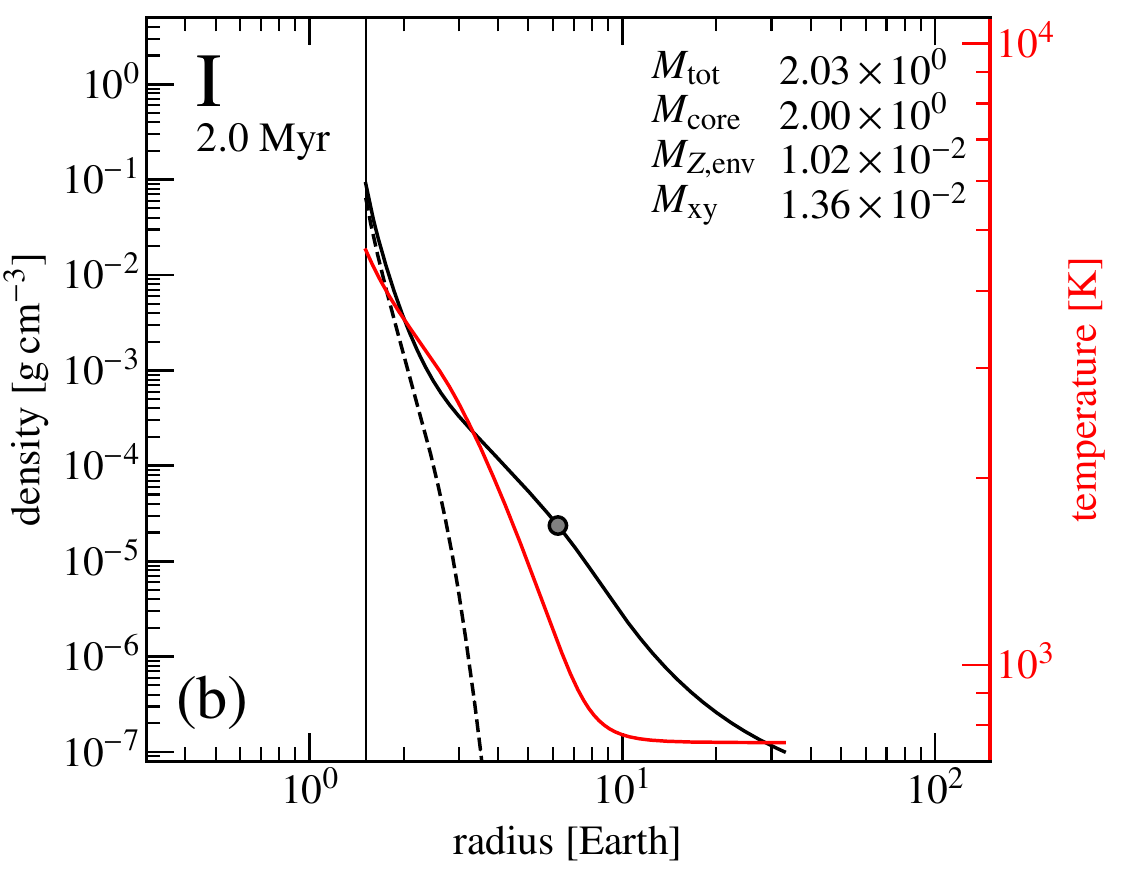}
     \includegraphics[width=0.333\textwidth]{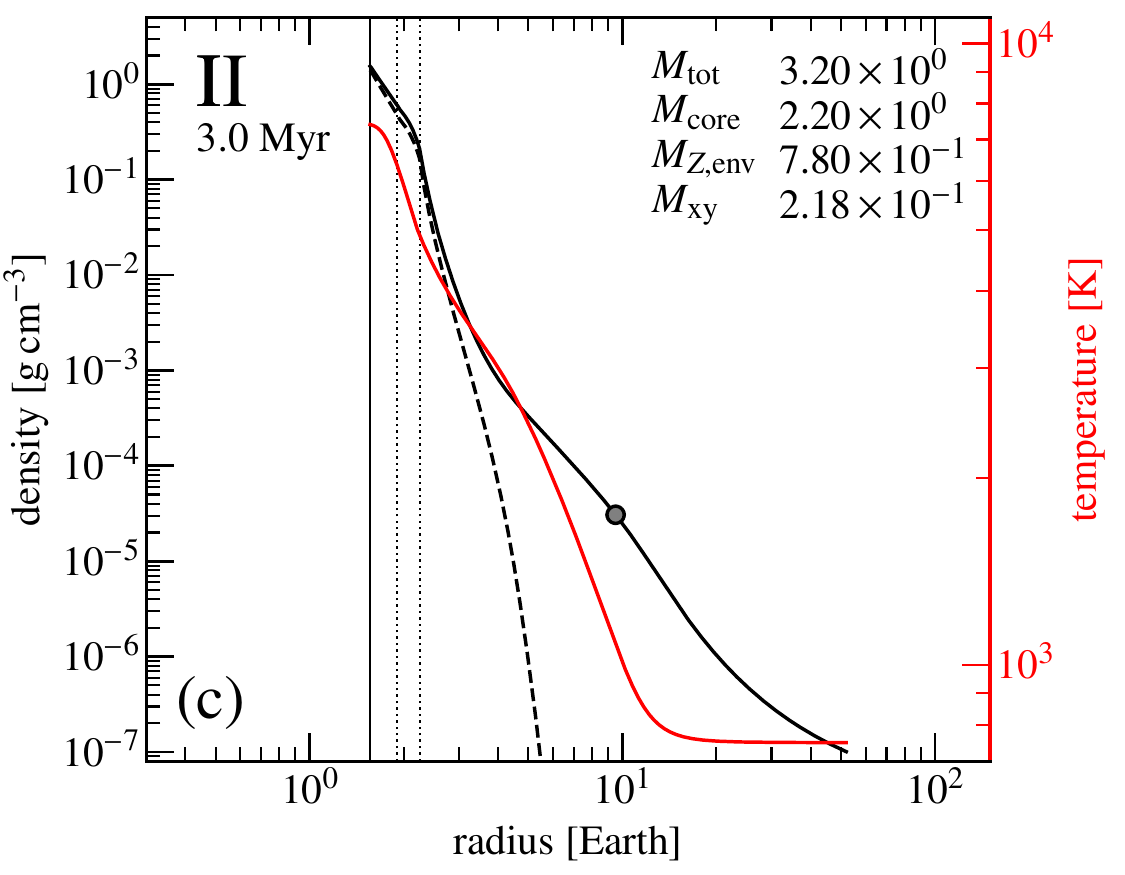}
     \includegraphics[width=0.333\textwidth]{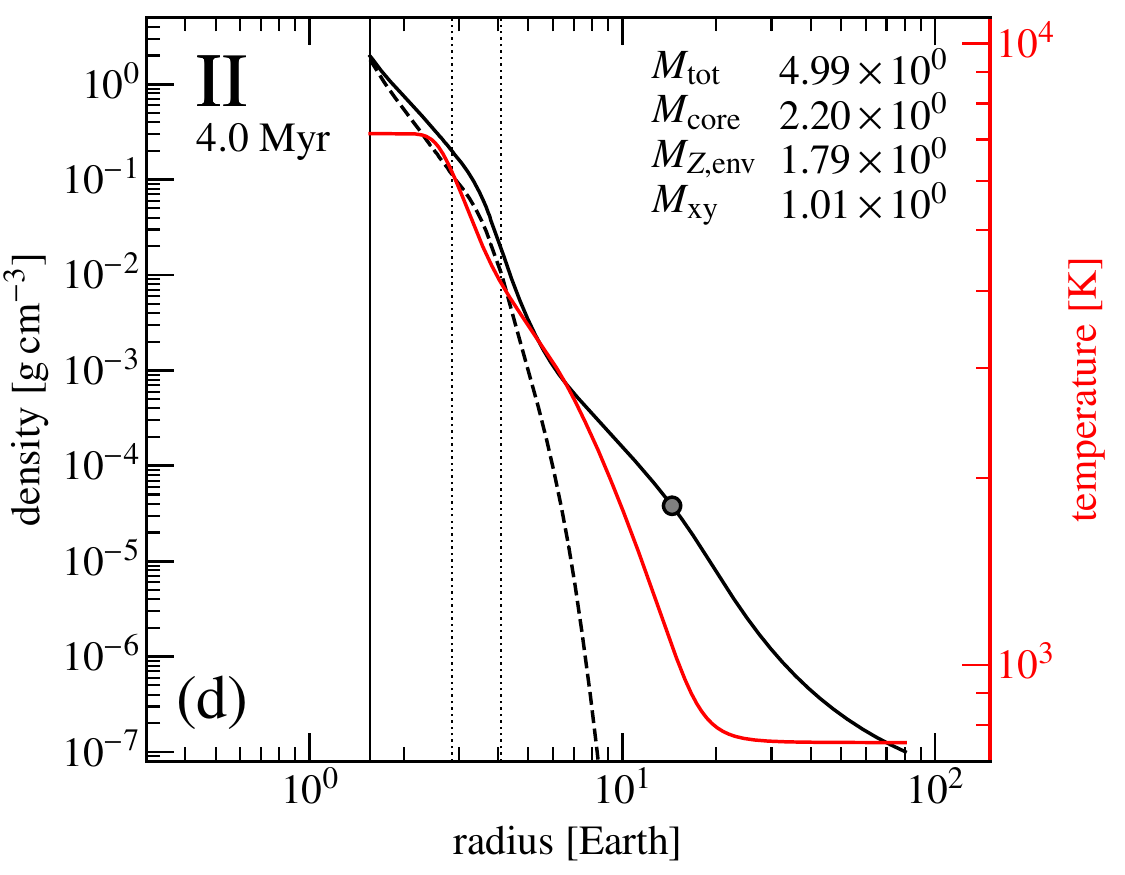}
     \includegraphics[width=0.333\textwidth]{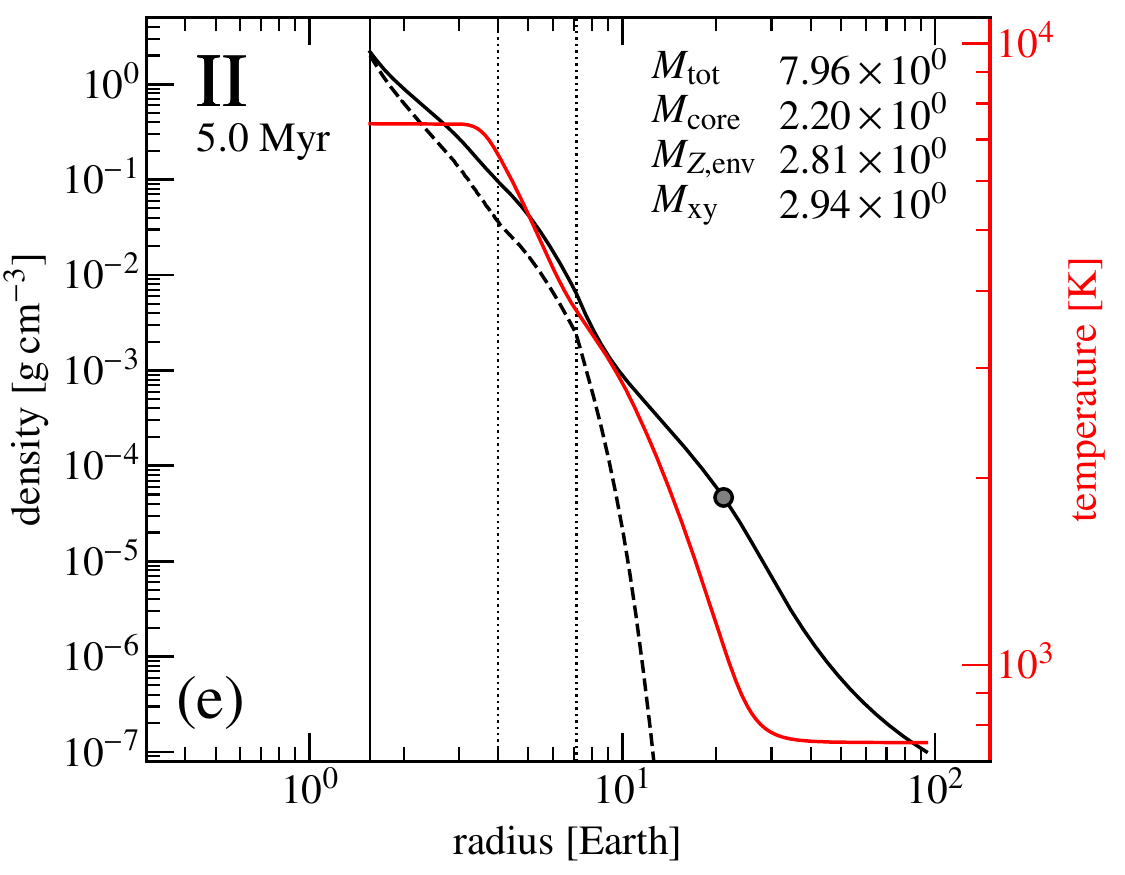}
     \includegraphics[width=0.333\textwidth]{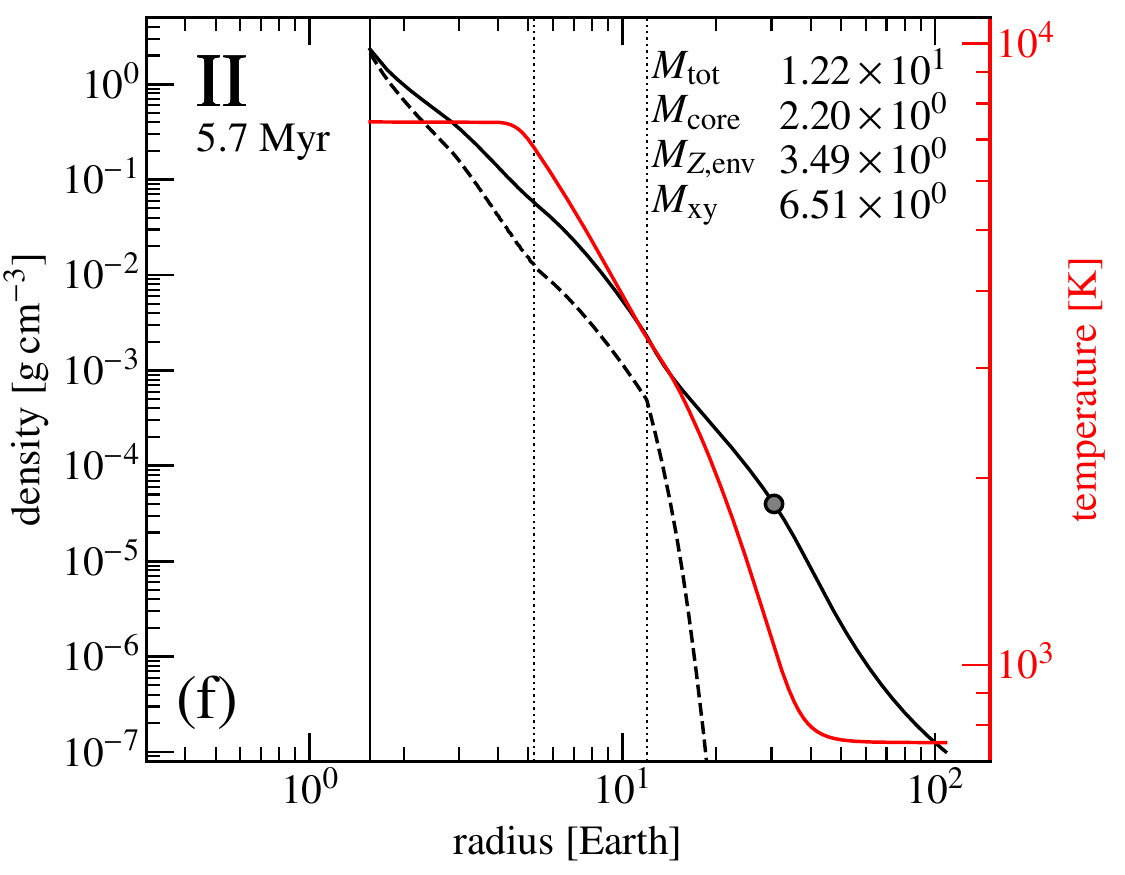}
     \caption{Evolution of envelope profiles of temperature (red), total density (solid black) and vapor density (dashed solid) for the standard model (0.2 au, $\dot{M}_\mathrm{peb}=10^{-6}\,M_\oplus\,\mathrm{yr}$, FEOS with \texttt{Kraus} liquid-vapor curve, molecular opacities). The core-envelope boundary (CEB) is indicated by the vertical solid line, while the vertical dotted lines denotes the boundaries between the frozen zone (region 2b), the well-mixed region 2a, and the saturated region 1. These boundaries expand with time, creating a gradient in the composition. The radiative-convective boundary is indicated by the grey dot. Towards the end the gas accretion picks up speed as the crossover mass is reached. The temporal evolution is available as an \textbf{online movie}.
     }
     \label{fig:standard}
\end{figure*}
The differential equations are solved with scientific python's (\texttt{scipy}) \texttt{bvp\_solve} routine. The solution technique is described in previous works \citep{AscherEtal1994,KierzenkaShampine2001,ShampineEtal2006}.  Based on an initial (or intermediate) guess for the structure, \texttt{bvp\_solve} will calculate the residual of the numerical derivative (from the guess structure) and that of the ODE expressions of \eq{ss-all} for all grid points and on the boundary. \texttt{bvp\_solve} will then adjust the solution until the residuals have become small enough. In our code, the solution structure to previous time is also available, which is needed to calculate the luminosity (see \eq{Lint}). For example,
\begin{equation}
    \frac{dE}{dt} \rightarrow \frac{E(m)-E'(m)}{\Delta t}
\end{equation}
where primes denote the previous time. The energy at the previous time $t'$ but the present mass grid $m$, $E'(m)$, is obtained by interpolation. Note the implicit formulation; the luminosity $L$ depends on the new structure.

At $t=0$ the simulations are initialized with a core mass small enough to render the envelope close to isothermal, with which the analytic (isothermal) guess solution will be close to the (known) analytical solution for constant $T$. The precise initial value is immaterial to the further evolution. At all other times, we start from the previous solution structure or we predict the solution structure at time $t$ by extrapolating the previous results.


\section{Results}
\label{sec:results}
We consider three disk locations: 0.2, 1, and 5 au and four different pebble accretion rates: $10^{-4}, 10^{-5}$, $10^{-6}$, and $10^{-7}$ $M_\oplus\,\mathrm{yr}^{-1}$. The pebble radius, which in our model setup only affects the opacity, is varied between "None" (no additional opacity from grains) and 100 $\mu$m. Other parameters are listed in \Tb{pars}. For simplicity, it is assumed that pebbles consist of silicates. At 5\,au, where the ambient disk temperature is only 150\,K, it is natural to expect that pebbles also contain \hho ice, which would ablate from the pebbles at much lower temperatures than \sioo. Therefore, our results for 5\,au primarily serve to illuminate trends upon varying the physical conditions, rather than being an attempt to accurately model a particular planet (e.g.\ Jupiter). 

Simulations are run until the crossover mass is reached. \Tb{results} presents the properties of the planet and atmosphere at this point for all model runs. It lists the core mass, vapor mass of the envelope $M_\mathrm{env,Z}$, the hydrogen and helium mass in the envelope $M_\mathrm{xy}$, the time after which crossover is reached $t_\mathrm{cross}$, the corresponding timescale for gas accretion ($t_\mathrm{xy} = M_\mathrm{xy}/\dot{M}_\mathrm{xy}$, and the extent of the region where metals constitute more than 5\% by mass $r_\mathrm{0.05}$. We also provide the average metallicity $Z_\mathrm{dc}$ within this region (including the core), which may be considered a measure for the diluteness of the heavy metals in the central regions (dilute core). Hence, for a sharp transition between a $Z=1$ core and a metal-poor envelope $Z_\mathrm{dc}=1$. But in cases where vapor and hydrogen/helium mixes, $Z_\mathrm{dc}$ decreases. In the limit that the central region is of homogeneous composition of $Z^\ast>0.05$, $Z_\mathrm{dc}=Z^\ast$.

\subsection{Baseline model}
In \fg{standard} the density and temperature profiles of the standard model (conditions applicable to 0.2 au, real EoS with the \citealt{KrausEtal2012} liquid-vapor curve, $\dot{M}_\mathrm{Peb}=10^{-6}\,M_\oplus\,\mathrm{yr}^{-1}$, no opacity from pebbles) are provided in an evolutionary sequence. Snapshots are shown every million year and at the end when the critical metal mass is reached. Initially ($t\lesssim1$ Myr) the vast majority of the accreted pebbles settle to the core without significant loss to thermal ablation, which is the classical situation. The envelope can be divided into a convective interior region and a radiatively-supported outer region, characterized by constant $T$. Nonetheless, because of the increasing temperatures near the base of the envelope, an increasingly larger fraction of their mass will be lost to the envelope by ablation. This is already clear after 1 Myr (\fg{standard}a) from the very steep increase of the vapor density, which follows the saturation density (Phase I). 

Since the saturation density is an exponential function of temperature, the regions close to the CEB become dominated by vapor. Because of the much higher molecular weight of the \sioo vapor, both density and temperature shoot up, as we explained in Paper I. Still, as of 2 Myr (\fg{standard}b) pebbles are only partially ablated, because the envelope's ability to absorb is limited. This absorption increases with time, however, and a point will be reached where the pebbles sublimate fully. This marks the end of Phase I (``direct core growth''). Compared to Paper I we note that the transition core mass we obtain here is higher by about a factor of two, primarily because the wet adiabatic gradient we employ in this work renders this region less hot.

\begin{figure*}[tb]
    \sidecaption
     \includegraphics[width=0.7\textwidth]{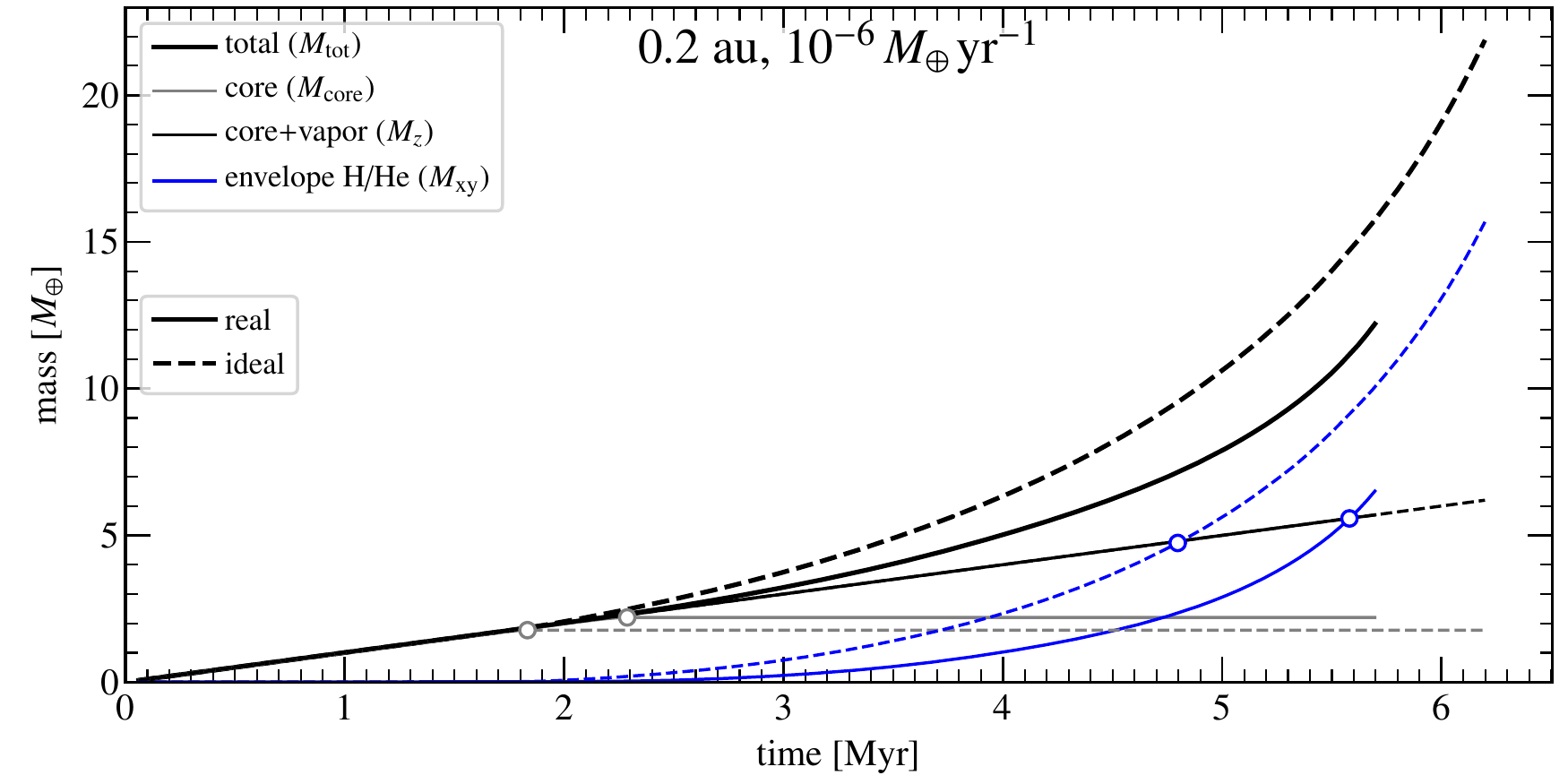}
     \caption{Comparison of time-evolution between the real EoS (solid lines) and ideal EoS (dashed) standard model run (0.2 au, $\dot{M}_\mathrm{peb}=10^{-6}\,M_\oplus\,\mathrm{yr}^{-1}$, no opacity from pebbles). Masses (total, core, metal, and hydrogen+helium) are shown on the left y-axis and the temperature at the core-envelope boundary is plotted on the right y-axis. The points where the core mass is fixed (evolution to Phase II) are indicated by the open gray circle, while the points of crossover are indicated by the blue circles.}
     \label{fig:evol0}
\end{figure*}
\begin{figure*}[tb]
    \sidecaption
     \includegraphics[width=0.7\textwidth]{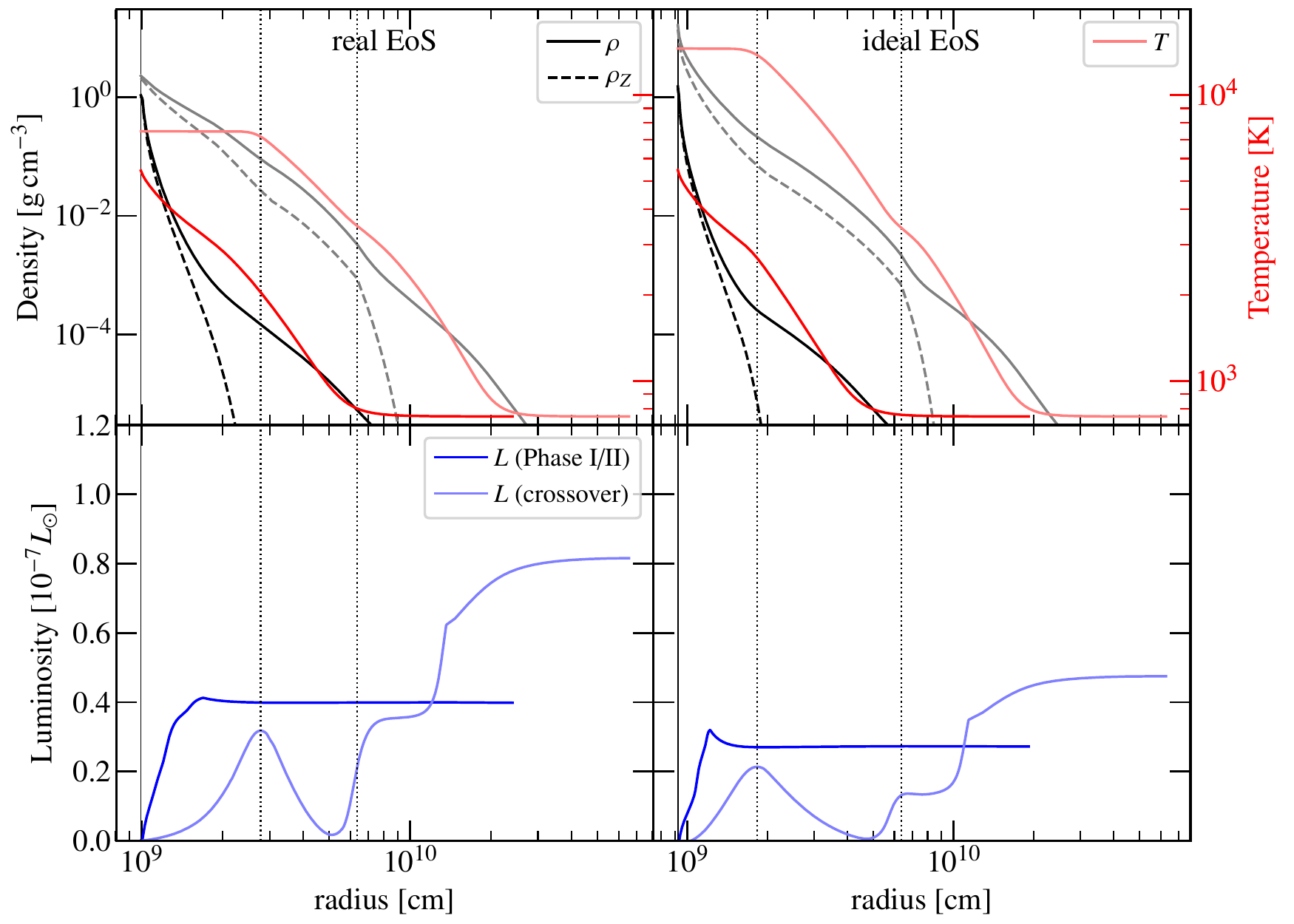}
     \caption{Profiles of density, temperature, and luminosity for the simulation with the real EoS (left panels) and the ideal EoS (right panels). Parameters as in \fg{evol0}. Profiles are plotted just after Phase II is reached (bright curves) and after crossover (dim curves). For the state at crossover, the vertical dotted line indicates the dividing line between the isothermal region 2b, the mixed region 2a, and the saturated region 1.}
     \label{fig:analysis}
\end{figure*}
In Phase II, the inner envelope is characterized by two new regions: the isothermal, compositionally-"frozen" zone (2b) and a convective region (2a) of constant metallicity $Z_\mathrm{cnv}=\rho_Z/\rho$.  The frozen zone emerges because there is not enough energy present to effectively mix this region. Pebbles, no longer reaching the core, deposit their energy in higher layers. Because of the rising temperatures, the interface between regions 1 and 2a moves out (in mass space), causing the metallicity of region 2a to drop as region 1 has the lower $Z$ even though it is saturated. For region 2, most of its energy arises from the compression of the envelope. Nevertheless, it is insufficient to mix the entire region 2 to uniform $Z$; only the upper layers (region 2a) do. The boundary between regions 2a and 2b is indicated by the thin vertical solid line in \fg{standard}. With time this line moves outward: the frozen region (2b) becomes larger. At the same time, as more metal-free nebular gas enters the atmosphere, the metalliticy of the well-mixed region (2a) drops. The composition at a certain mass shell in region 2b is set at the time when this mass shell transferred from 2a to 2b (crossed $r_\mathrm{iso}$), being frozen ever since. As a result, a compositional gradient develops in the frozen region (2b).

As all accreted solids are turned into vapor, the total amount of high-Z vapor in the envelope steadily increases. By 4\,Myr, (\fg{standard}d) it has exceeded the core mass.  Initially (\fg{standard}c-d), the envelope is vapor-dominated ($M_\mathrm{Z,env} \gg M_\mathrm{xy}$). Nevertheless, accretion of H/He gas slowly catches up with the (constant) accretion rate of pebbles. After 5 Myr (\fg{standard}e) H/He gas is accreted at a higher rate than pebbles and after 5.4 Myr (\fg{standard}f) the H/He fraction has exceeded the combined mass of the core and the high-Z vapor -- our definition of crossover. We have verified that by crossover the planet accretes H/He gas on a timescale ($t_\mathrm{xy}=1.7\,\mathrm{Myr}$) shorter than its evolutionary time ($t_\mathrm{cross}=5.35\,\mathrm{Myr}$). The planet will therefore turn into a hot-Jupiter, provided the disk is able to keep supplying gas. Conversely, when the disk disperses before $t_\mathrm{cross}$ hot-Jupiter collapse is avoided and the planet ends up as a (mini-)Neptune planet.

\subsection{Ideal vs real EoS}
\label{sec:EoS-comp}
As many studies, including ours in Papers I and II, are based on an ideal EoS \citep[\eg][]{LeeEtal2014,GuptaSchlichting2019}, we pay some attention to compare our results obtained with the real EoS to the results based on the ideal EoS (\se{EoS-ideal}).  In \fg{evol0} we contrast the time-evolution of several key quantities between the non-ideal EoS runs, while in \fg{analysis} we contrast the density and temperature profiles at two points in their evolution: (i) when pebbles fully evaporate (transition Phase I and II); and (ii) at the crossover mass.

In Phase I the evolution of the ideal and non-ideal runs proceed similarly, because non-ideal effects are unimportant for low-density atmospheres. When a vapor-dominated inner region appears, the temperature and, especially, the density tend to rise steeply. However, the cusps that we see in this work are less pronounced than we saw before in Papers I and II. The reason is that the adiabatic gradient transitions to that of a wet adiabat, which drives $\nabla_\mathrm{ad}$ in both the ideal and real EoS to the same value. As can be seen from \fg{analysis} the profiles at the point of transition between Phase I and II are quite similar.

After reaching Phase II, there is no longer an energy source from accretion at the core-envelope boundary (CEB), which mitigates the temperature increase or even decreases it (real EoS) (see \fg{standard}). 
In the ideal case, the density near the CEB keeps increasing because the ideal vapor is still characterized by the same low $\gamma_\mathrm{ad}$, i.e., a highly compressible vapor. Densities then start to exceed the density of the core, which is unphysical. At the point of crossover the ideal run achieves densities of $\rho=15.5\,$g\,cm$^{-3}$ at the base of the envelope, while using the non-ideal EoS the core-envelope density is at $\rho=2.2\,\mathrm{g\,cm}^{-3}$ still less dense than the core. In contrast, in the non-ideal runs the vapor becomes supercritical and much harder to compress (see \fg{EoS-FEOS}). Because the vapor in the ideal runs is confined to a shell immediate to the CEB, the intake of H/He-rich (hydrogen and helium) material develops stronger in the ideal case with respect to the non-ideal EoS run and the crossover mass is somewhat smaller.

Barring the difference in the density and temperature profiles, which differ greatly in the central regions, there are no major qualitative differences between the outcome of the ideal and non-ideal runs  when one considers the integrated quantities at crossover (\ie $t_\mathrm{cross}$, $M_\mathrm{core}$, $M_\mathrm{cross}$;  
see \Tb{results}). Rather than the equation of state, uncertainties pertaining the liquid-vapor curve, (multi-species) chemistry and dissociation (neglected here) will have far greater effects. Therefore, simulations based on the ideal EoS provide acceptable, first order results.  Still, the distribution of (vapor) density and temperature is of importance when one considers the long-term evolution of the interior structure and the potential for mixing.  

\fg{analysis} also illustrates how our implementation regarding the avoidance of negative luminosity works out. The vertical line indicates the dividing line between the isothermal region (2b) where the composition is frozen and the region (2a) where convection operates to mix the vapor and hydrogen/helium gas to uniform $Z$. This distinction is reflected in the luminosity profiles. In the frozen region $PdV$ compression liberates energy and the luminosity increases. Conversely, in region 2a mixing consumes energy and $L$ decreases. The luminosity hence peaks locally (by construction) at the interface of regions 2a and 2b.

\begin{figure}[t]
     \includegraphics[width=\columnwidth]{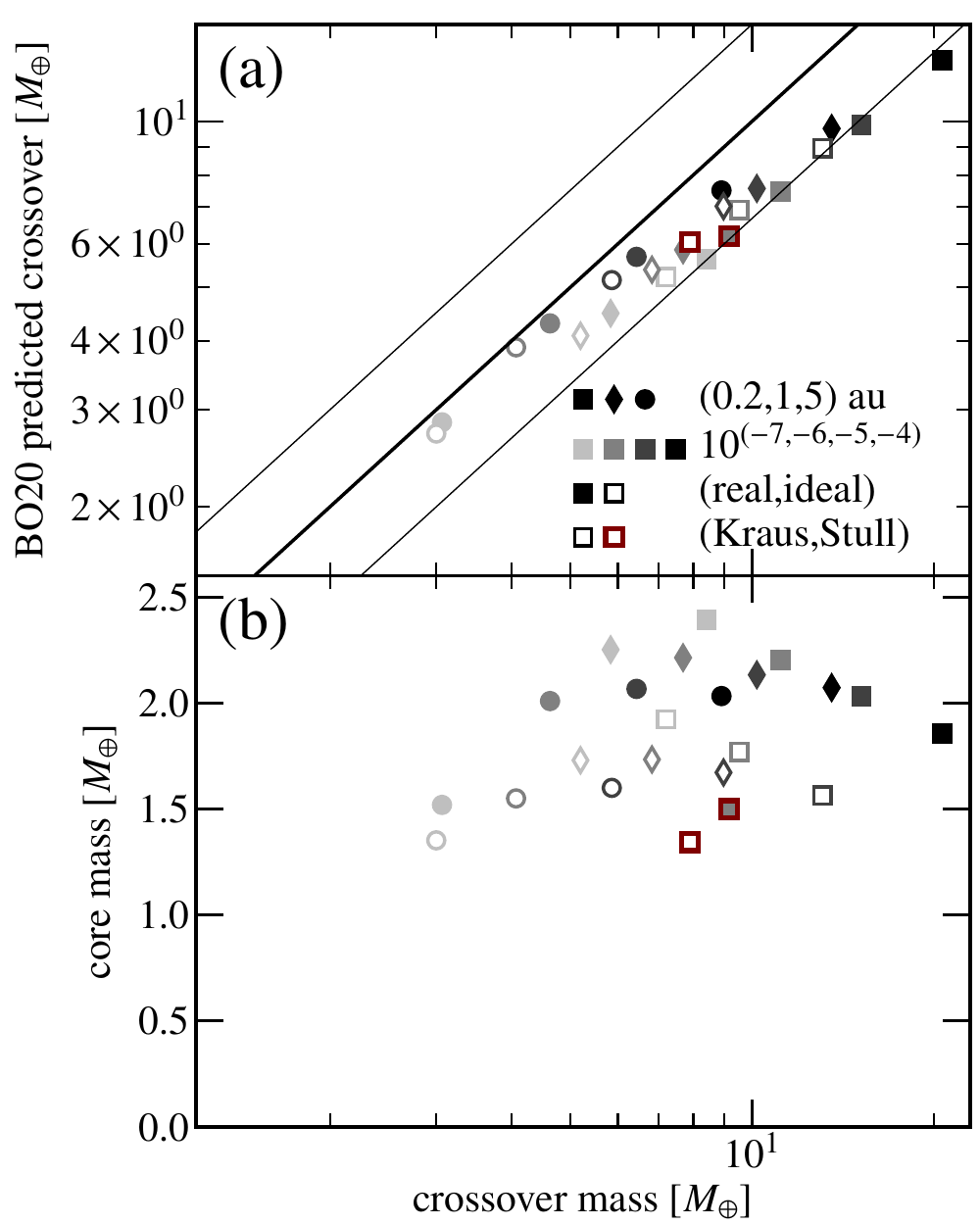}
     \caption{(a) Analytically-predicted (Paper II) vs numerically-obtained crossover mass for simulations at different distances (symbols), pebble accretion rates (symbol edge color), and equation of states (filling). The thick diagonal line indicates a perfect match (predicted=numerical) while the thin lines indicated a 50\% offset. (b) the core masses of the simulation runs. Square symbols with a brown edge color employ the \citet{Stull1947} liquid-vapor curve.}
     \label{fig:corecross}
\end{figure}
\begin{figure*}[tb]
    \sidecaption
     \includegraphics[width=0.7\textwidth]{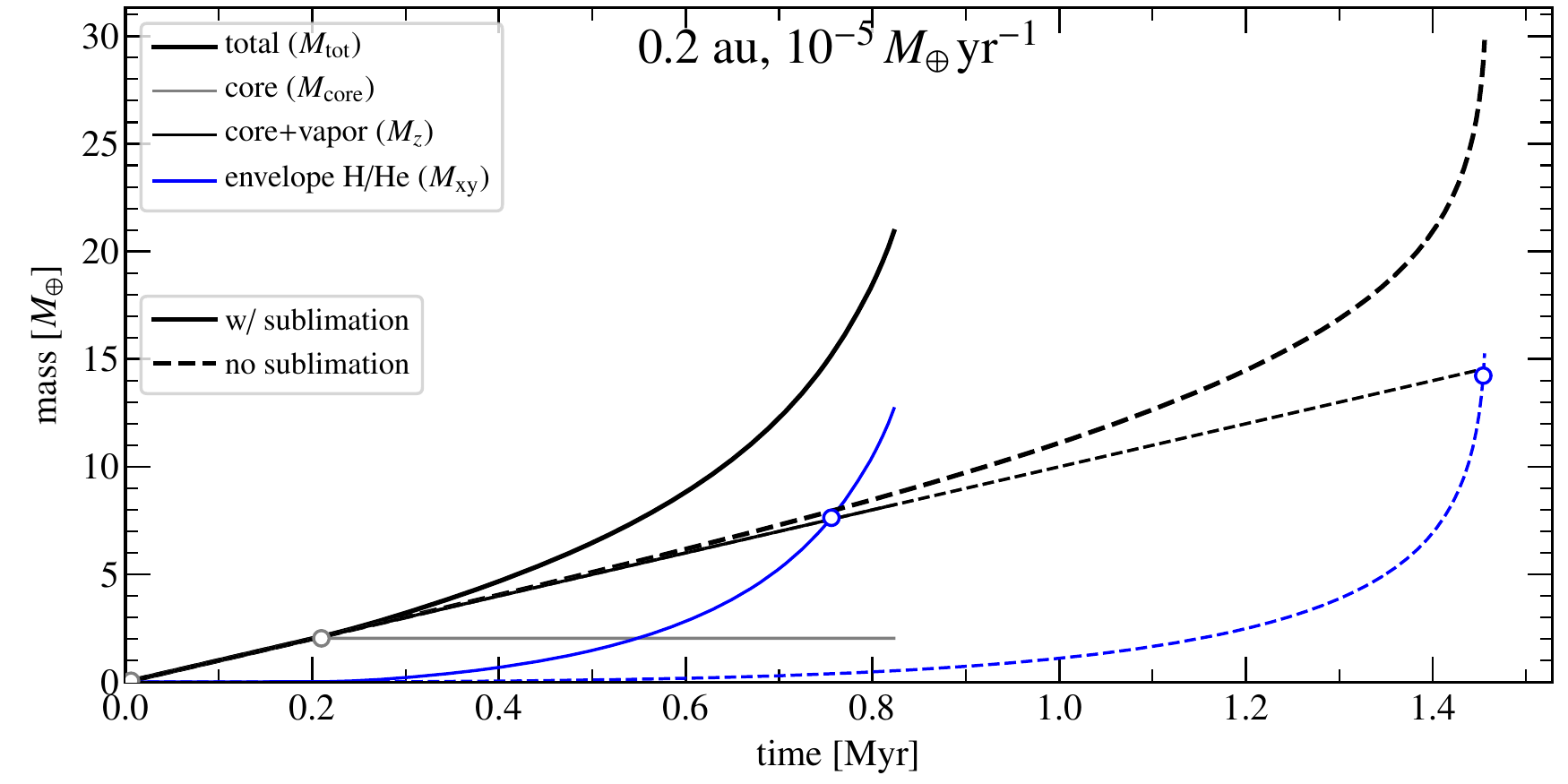}
     \caption{Temporal evolution of two runs with our (standard) choice of evaporation (solid) and for the classical choice where solids end up in the core (dashed). Other parameter are the same (\eg $\dot{M}_Z=10^{-5}\,M_\oplus\,\mathrm{yr}^{-1}$). Due to the absorption of vapor by the envelope and the concomitant reduction of the local pressure scaleheight, the intake of nebular gas ($M_\mathrm{xy}$) speeds up in the run where pebbles sublimate.}
     \label{fig:evol1}
\end{figure*}

\subsection{Comparison to analytical prediction}
In Paper II we derived analytical formulae for conditions at crossover based on a simple three-layer model of the atmosphere. Specifically, for the crossover mass we derived:
\begin{align}
    \nonumber
    M_\mathrm{cross} \approx\, & 8.0\,M_\oplus
    \left( \frac{\kappa_\mathrm{rcb}}{\mathrm{0.01\,g\,cm^{-3}}} \right)^\frac{1}{6}
    \left( \frac{d}{\mathrm{au}} \right)^\frac{7}{108}
    \left( \frac{T_\mathrm{vap}}{2500\,\mathrm{K}} \right)^\frac{8}{27} \\
    & \times  \left( \frac{\dot{M}_Z}{10^{-5}M_\oplus\,\mathrm{yr}^{-1}} \right)^\frac{1}{6}
    \left( \frac{M_c}{M_\oplus} \right)^\frac{1}{2}
    \label{eq:BO20eq30}
\end{align}
\citep[Eq.\,(30) of][the crossover mass is twice the critical metal mass]{BrouwersOrmel2020}
where $T_\mathrm{vap}$, a parameter in the three-layer model, identifies the outer boundary of the vapor-dominated zone and $\kappa_\mathrm{rcb}$ is the opacity at the radiative-convective boundary.

We test \eq{BO20eq30} in \fg{corecross}a, where we obtain $\kappa_\mathrm{rcb}$ directly from our simulations (so this value is different per run) while the evaporation temperature is fixed at its fiducial value of $2500\,\mathrm{K}$.. The analytical prediction follows the numerical results quite well, especially given the assumptions that our analytical model contains (\eg a fixed adiabatic index).  Nevertheless, the deviation between the numerical and analytical results do not deviate much.  Even though the non-ideal runs result in higher core masses, the $M_\mathrm{core}$--$M_\mathrm{cross}$ relation is not affected by the choice of the EoS. This is because the relation reflects a more basic principle: core and atmosphere mass are coupled by the thermal state of the envelope. For example, a hotter envelope results in more evaporation,  a smaller core mass, which gives rise to a smaller crossover mass.  We attribute the lower predicted crossover mass of the 0.2 au runs to the breakdown of the isothermal assumption on which \eq{BO20eq30} relies in the outer radiative region. This implies that the atmosphere mass (for a given core mass) is lower; hence a higher crossover mass. 

Compared to our previous works (Papers I and II) core masses are higher, due to the inclusion of a wet adiabat, as discussed above.  A second reason for the higher core masses is that we are now using the \citet{KrausEtal2012} liquid-vapor curve, which lies below the \citet{Stull1947} curve that we used before (see \fg{EoS-FEOS}); for a given temperature vapor will become saturated at a lower density for the Kraus LV-curve. As a result, the value of the core mass after Phase I is typically around 1.5--2\,$M_\oplus$. In \fg{corecross}b it can also be seen that the core masses obtained from the real EoS simulations (filled symbols) are slightly higher than those obtained from the ideal EoS simulations.
The reason for this effect is that the high-density, high-temperature cusps that develop towards the end of Phase I, although the effect is suppressed by the wet adiabat, are still somewhat more pronounced in the ideal simulations, especially in terms of density (\fg{analysis}) due to the fact that the non-ideal vapor is less compressible.\footnote{When we would remove the wet adiabatic from the numerical model the discrepancy between the ideal and non-ideal runs would magnify: the non-ideal EoS runs would still result in $\approx$1.5--2\,$M_\oplus$ cores, but the core mass in the ideal runs reduces to $\approx$1\,$M_\oplus$. The difference can be entirely attributed to the compressible nature of the ideal gas, exacerbating the cusps of high temperature and density.} The dependence of the core mass on the pebble accretion rate is ambiguous: at 5 au there is a positive trend, while at 0.2 au the trend is negative. Competing effects are at play that make this trend rather shallow. A higher $\dot{M}_\mathrm{peb}$ tends to saturate the envelope, prolonging Phase I, but it also raises the temperature, allowing more vapor to be absorbed. 

\subsection{Benchmark test}
To put these results in context, we comparison our results to the classical assumption where pebbles are immune to sublimation and end up to the core. In \fg{evol1} this comparison is being made for models run at 0.2 au and for $\dot{M}_Z=10^{-5}\,M_\oplus\,\mathrm{yr}^{-1}$. Since in the classical model there is no vapor in the envelope, the line "core" and "core+vapor" overlap. In the classical run the core mass eventually exceeds 10 $M_\oplus$, whereas in the case with pebble sublimation it reaches just over 2 $M_\oplus$. Despite the relatively small core, the run allowing for pebble sublimation takes in a higher fraction of nebular gas, as illustrated by the slope of the blue lines (which gives $\dot{M}_\mathrm{xy}$). The reason is that (part of) the vapor mixes with the nebular gas (in region 2a) to lift the molecular weight of these parts of the envelope, reducing the local pressure scaleheight. To satisfy pressure balance, more nebular gas is accreted. 

\fg{benchmark} presents a more complete study, also accounting for how the results are affected by entropy advection. In this scatter plot, we show the crossover mass of 27 runs as function of distance and pebble accretion rate (x-axis). The color fill indicates how polluted the envelopes are, ranging from non-polluted (white; by definition the runs without sublimation have $M_\mathrm{Z,env}=0$) to polluted (black).  Furthermore, the symbol edge color gives an indicating of the state of runaway gas accretion at the time of crossover.  It gives the timescale of hydrogen/helium accretion ($t_\mathrm{xy}$) over the time at crossover $t_\mathrm{cross}$ with light colors signifying this ratio is low, indicating rapid gas accretion, and dark colors that the ratio is about unity (runaway accretion has not been reached yet). See \Tb{results} for the numbers.

In the case without sublimation (circles) there is by definition no Phase II and there is also no isothermal layer; the continuous accretion of pebbles onto the cores will render the luminosity positive everywhere. In terms of the critical metal mass, however, the outcome is similar to our standard runs, especially in the outer disk and for low accretion rate. This is unsurprising, because the combination of low opacities and low $\dot{M}_\mathrm{peb}$ render the envelope so cold that pebbles hardly sublimate! (See the last two lines of \Tb{results} and note the low $M_\mathrm{Z,env}$). On the other hand, 
in the inner disk runs, where pebbles do vaporize, the vapor and associated higher-molecular weight of the atmosphere accelerates the intake of H/He gas, because the mixing results in a more compressed envelope (Paper II).

At 5 au, we can compare our results to those of \citet{HoriIkoma2010}, who consider almost the exact similar parameters in their study. The critical core masses that they obtain are approximately, 2, 3, and 5 $M_\oplus$, respectively, for the three accretion rates. These numbers are slightly higher than ours, but not significantly so. The critical core mass in their study may be slightly higher than half the crossover mass, as it is set by the failure of hydrostatic balance. More generally, these results are rather sensitive to the opacity model, for which we have used simple power-laws.\footnote{Another choice is to use the \citet{BellLin1994} opacity fit, as we did in Paper I and II. However, this results in even smaller values compared to \citet{HoriIkoma2010}.}

In the runs with entropy advection (triangles), the atmosphere is 50\% smaller in radius (see \se{bc-sol}). 
The layer between $R_\mathrm{adv}$ and $R_\mathrm{disk}$ is assumed to have the same entropy as the disk (the layer is not included in the envelope mass) and therefore the density at $R_\mathrm{adv}$ will be lower and the temperature higher.
Consequently, the atmospheres tend to have less H/He for a given metal mass, resulting in a higher crossover mass. Another feature is that gas accretion rates at crossover are not very high, $t_\mathrm{xy} \sim t_\mathrm{cross}$, especially for the inner regions. This suggests that these atmospheres have not yet entered runaway gas accretion. Otherwise, for all other runs, the critical metal mass always signifies the onset of runaway gas accretion.

\begin{figure}[t]
     \includegraphics[width=\columnwidth]{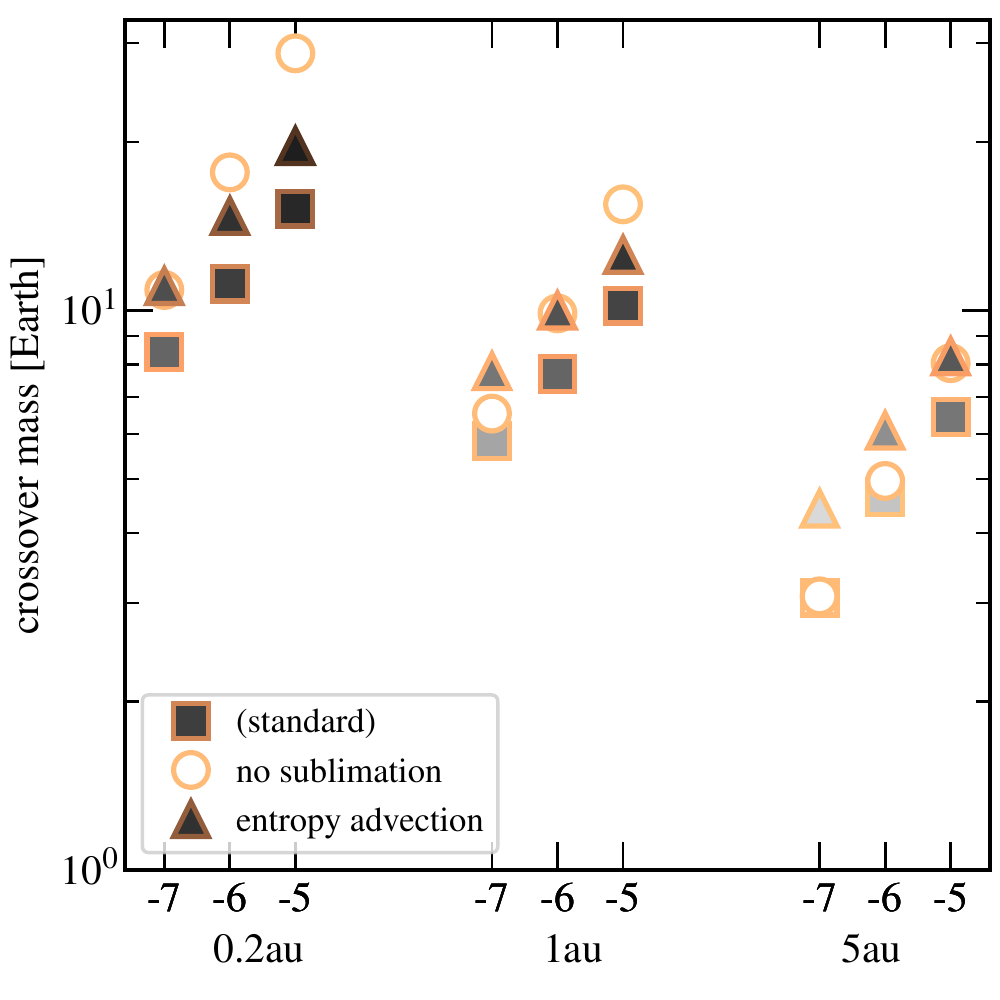}
     \caption{Comparison of our standard model setup (squares) with other model designs: sublimation-free runs (all solids reach the core; circles) and runs with entropy advection (triangles) as function of distance and pebble accretion rate (x-axis). The symbol filling indicates the level of pollution of the atmospheres at the point of crossover (white: no vapor; black: 50\% vapor). The symbol border color indicates how quick nebular gas is being accreted (black: $t_\mathrm{xy}/t_\mathrm{cross} \ge 1$; copper: $t_\mathrm{xy}/t_\mathrm{cross}=0$). See \Tb{results} for the precise values.}
     \label{fig:benchmark}
\end{figure}
\begin{figure}[t]
     \includegraphics[width=\columnwidth]{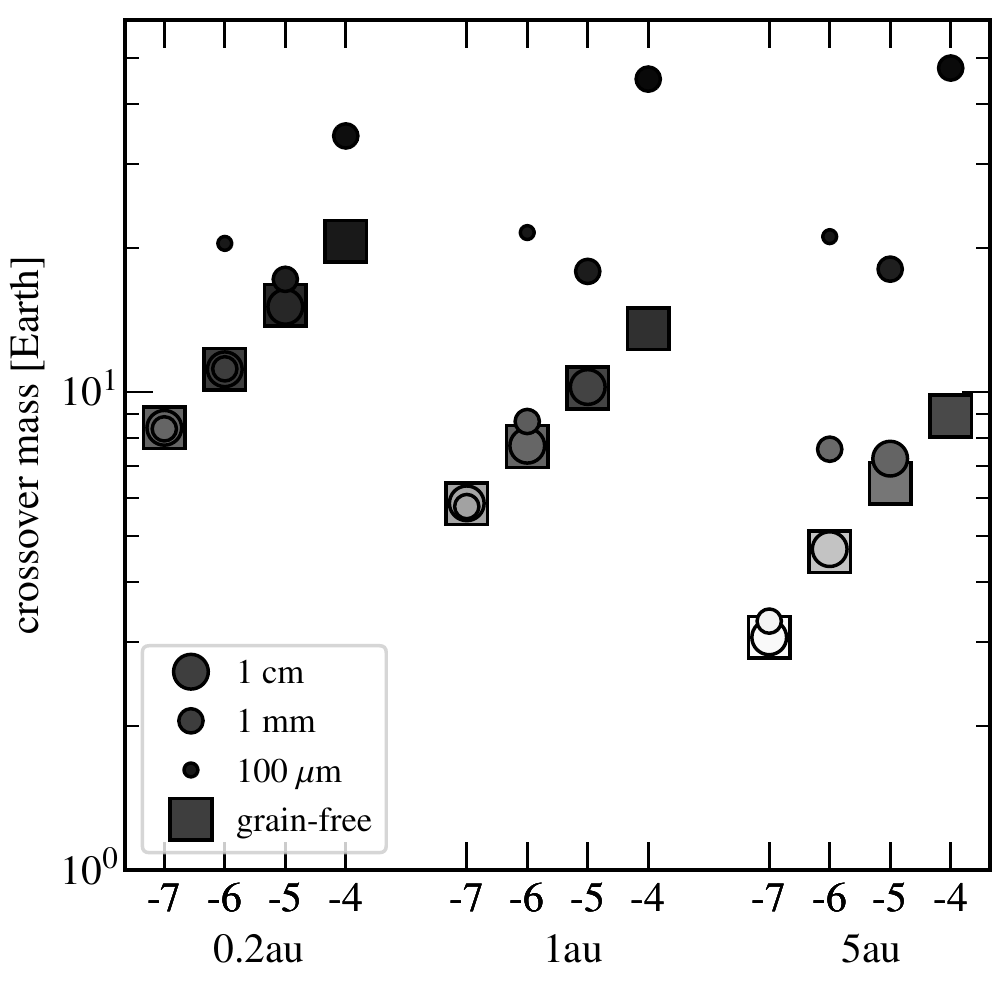}
     \caption{Dependence of crossover mass on pebble size. A different pebble size changes the opacity. "Grain-free" indicates that pebbles do not contribute. The symbol filling indicates the level of enrichment of the atmosphere at crossover (white: no vapor; black: 50\% vapor).}
     \label{fig:pebble-opacity}
\end{figure}
\subsection{Opacity of pebbles}
Until now we have assumed a grain-free atmosphere. This assumption is based on the belief that any grains would co-accrete with the gas, would readily coagulate and be removed by settling \citep{Mordasini2014,Ormel2014}. In addition, one can argue that much of the ``original'' grains have been converted into pebbles, planetesimals, or planets, rendering the abundance (much) lower than the ISM value ($\sim$$10^{-2}$) and also highly uncertain. Conversely, pebbles are accreted at a rate $\dot{M}_\mathrm{peb}$, in input parameter that can be straightforwardly converted into an opacity by \eq{kappa-peb} once we specify a pebble size. This value is then added to the molecular opacity.

\Fg{pebble-opacity} presents the crossover mass as function of pebble size (symbol), distance (x-axis) and accretion rate (x-axis). Large pebbles, simulations conducted in the inner disk regions, and simulations at low $\dot{M}_\mathrm{peb}$ do not deviate much from the grain-free result (squares; grain-free meaning no opacity from solids). In the inner regions, the molecular opacity is much higher and the pebbles contributes little. Conversely, in the outer disk a smaller pebble size and a higher pebble accretion rate enhance the pebble opacity, rendering atmospheres hotter and less dense. When pebbles dominate the opacity, we find that the crossover mass becomes much higher. For example, at 5 au for $\dot{M}_\mathrm{peb}=10^{-6}\,M_\oplus\,\mathrm{yr}^{-1}$, the crossover mass jumps from $\approx$3$M_\oplus$ (grain free/1 cm size pebbles) to $\approx$$7\,M_\oplus$ (1 mm pebbles) to $22\,M_\oplus$ (100\,$\mu$m pebbles). A similar effect is seen for the accretion rate: at 5\,au for 1\,mm pebbles the crossover mass scales as $M_\mathrm{cross}\propto\dot{M}_\mathrm{peb}^{0.4...0.5}$. When pebbles dominate the opacity, furthermore, the crossover mass tends to become independent of distance. The key reason is that $\kappa_\mathrm{peb}$  is independent of gas density: $r^2 v_\mathrm{sed}= GM t_\mathrm{stop}$ (force balance) and $r^2 v_\mathrm{sed} \rho_\mathrm{gas} s_\mathrm{peb}$ is therefore approximately constant in the Epstein limit (see \eq{kappa-peb}).

\begin{figure}[t]
    \includegraphics[width=\columnwidth]{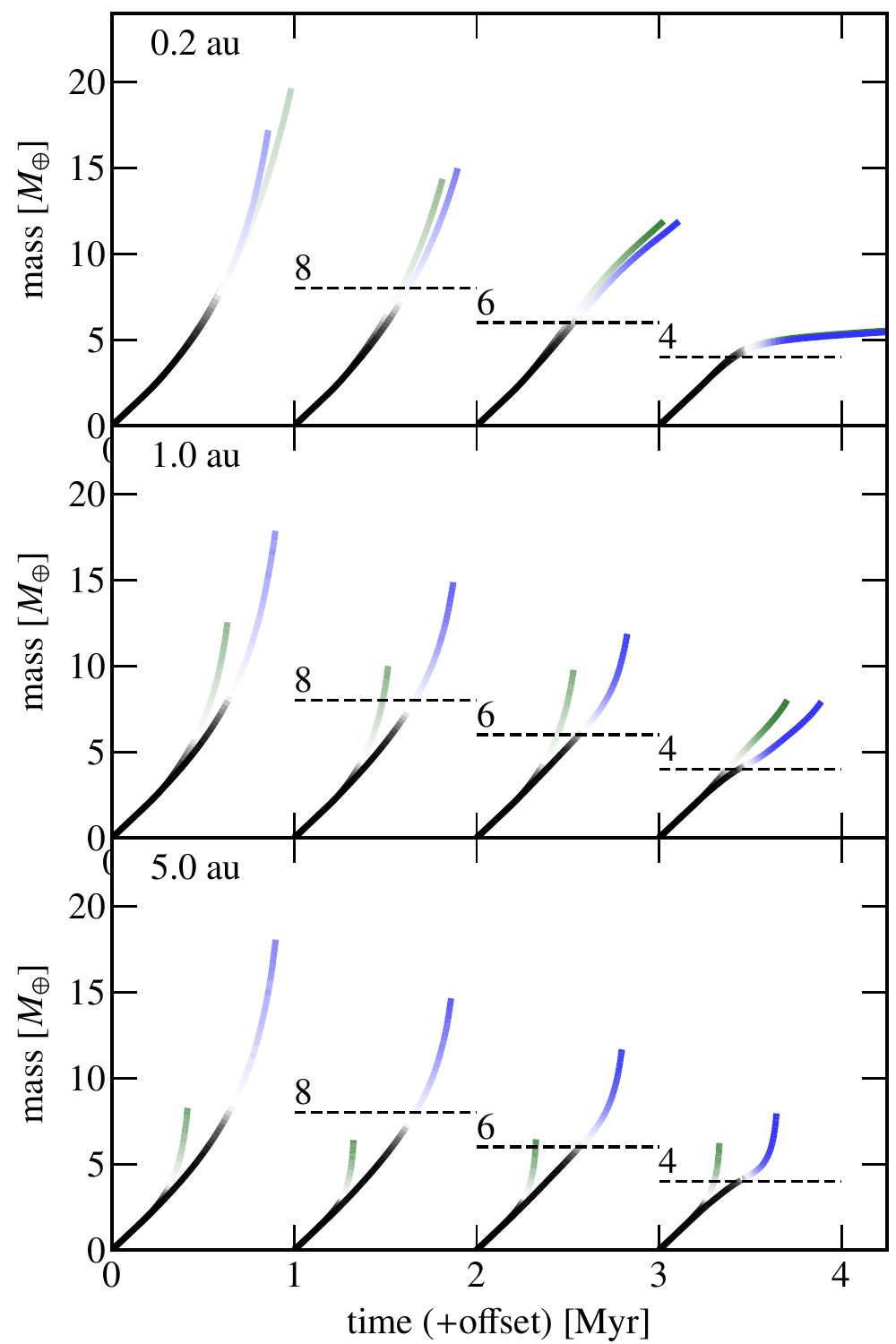}
    \caption{Simulations where the total amount of pebbles is limited to (from left to right): $M_\mathrm{Z,final} = \infty$ (unlimited; left), 8, 6, and, 4 $M_\oplus$ (right). Plotted is the total planet mass against time. Simulations terminate at crossover. Blue curves correspond to runs with additional opacity from mm-sized pebbles (the pebble opacity vanishes along with pebble influx). Green curves correspond to runs with only molecular opacities. The color indicates the nature of the accretion material, ranging from dominated by pebbles (black), nebular gas (blue or green) or similar (white).}
    \label{fig:phase3}
\end{figure}
\begin{figure*}[t]
    \sidecaption
    \includegraphics[width=0.7\textwidth]{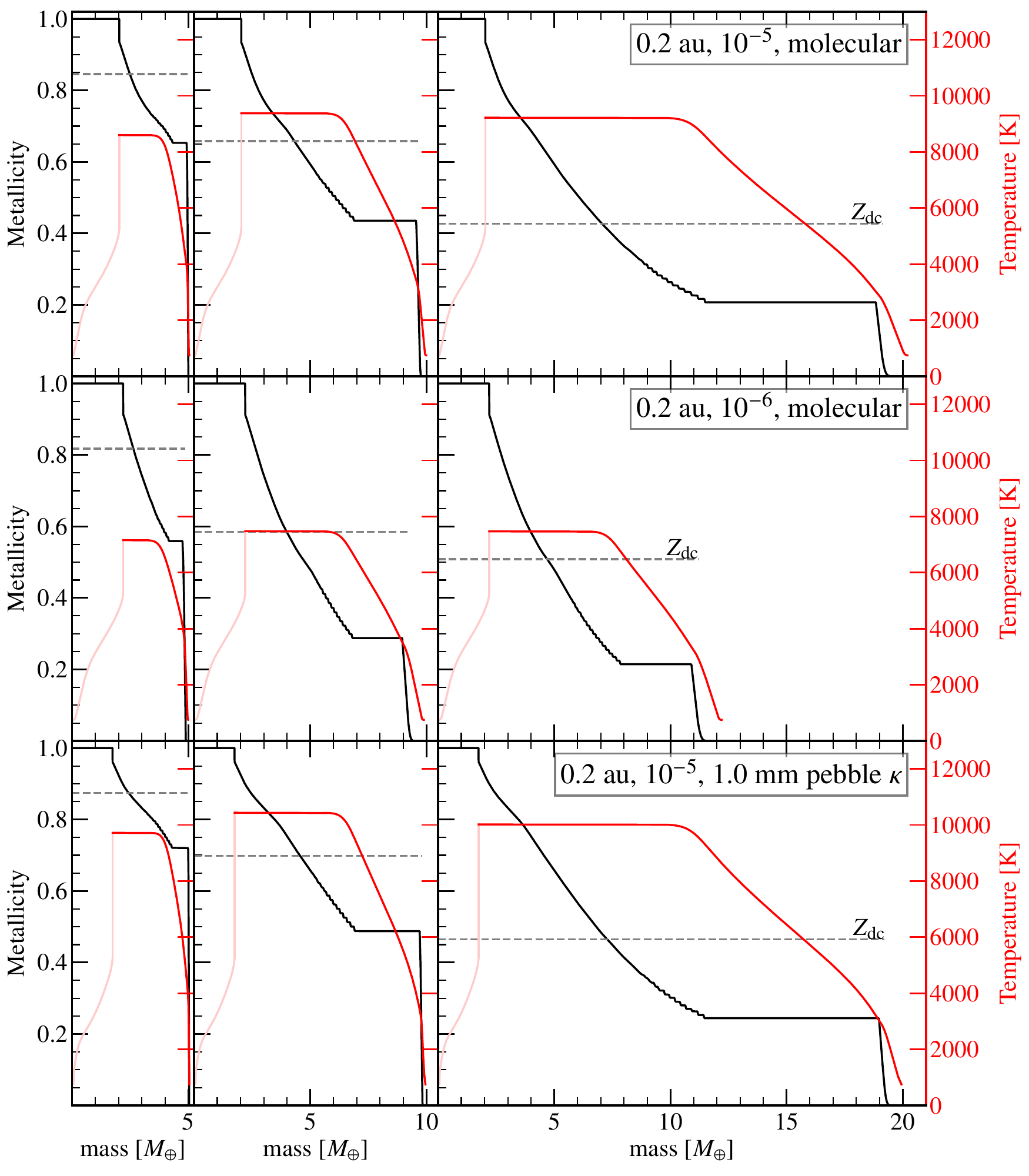}
    \caption{Metallicity and temperature profiles for several runs at 0.2 au: with an accretion rate of $\dot{M}_\mathrm{peb}=10^{-5}\,M_\oplus\,\mathrm{yr}^{-1}$ (top); a lower accretion rate of $\dot{M}_\mathrm{peb}=10^{-6}\,M_\oplus\,\mathrm{yr}^{-1}$ (center); and a pebble opacity corresponding to mm-sized pebbles. Profiles are shown when $M=5, 10$ and 20 $M_\oplus$ are reached (unless crossover is reached sooner). The value of the diluted core  $Z_\mathrm{dc}$ -- defined as the average over the region where $Z>0.05$ -- is also indicated. The temperature in the classical core region, which is not modelled, indicates the temperature at the core-envelope boundary when the material was incorporated.}
    \label{fig:profiles}
\end{figure*}
\subsection{Phase III}
In all preceding simulations we have assumed a continuous and constant flux of pebbles up to the crossover mass. In reality, it is conceivable that the pebble flux will stall, when, for example, the disk has run out of pebble-building material, an exterior giant planet forms stops the pebble flux interior to it, or when the planet itself has has reached either the pebble isolation mass \citep{AtaieeEtal2018,BitschEtal2018} or the flow isolation mass \citep[for small pebbles]{RosenthalMurray-Clay2020,KuwaharaKurokawa2020}. In these cases, the evolution is set by the self-contraction of the envelope (Phase III).  We therefore run a suite of simulations where we set the final solid mass $M_\mathrm{Z,final}$ at 4, 6, and 8 $M_\oplus$, respectively, as described in \se{phase3}.

\Fg{phase3} shows the total planet mass (metals and gas) as function of time for these runs, including the standard run with unlimited pebble supply ($M_\mathrm{z,final}=\infty$; left-most branches). The (initial) pebble accretion rate is $\dot{M}_\mathrm{peb}=10^{-5}\,M_\oplus\,\mathrm{yr}^{-1}$. Both runs, assuming grain-free envelopes (large pebbles; green) and pebble opacity runs with $s_\mathrm{peb}=1$ mm (blue), are shown. 

As discussed in the previous sections, runs with an additional opacity contribution from pebbles reach higher crossover masses. However, when the pebble accretion shuts off after $M\approx M_\mathrm{Z,final}$ (dashed horizontal lines; the total planet mass is dominated by metals at this point), the envelopes will nonetheless become grain-free, greatly reducing the opacity, especially in the outer disk. Consequently, these envelopes start to cool more rapidly. Accretion of metals (black) gives way to accretion of hydrogen/helium gas (color). The reason is simply that the Kelvin-Helmholtz timescale is short for grain-free atmospheres. Only when pebble accretion is halted at $M_\mathrm{Z,final}=4M_\oplus$ at 0.2 au, do we observe a transition to a slower evolutionary growth. The reason is that $t_\mathrm{KH}$ is a strong function of both opacity as well as planet mass \citep{HoriIkoma2010,LeeEtal2014}.

While our numerical results for Phase II broadly agree with the analytical predictions of Paper II, this no longer applies to Phase III. In Paper II we assumed that the composition of the vapor-dominated part of the envelope stayed homogeneous. Therefore, we argued that with addition of H/He gas and subsequent reduction in the molecular weight, the gas provided more hydrostatic support. Consequently, our prediction was that gas accretion would slow. However, in reality there is not enough energy to fully mix the vapor-rich regions with the incoming H/He gas -- a caveat that we already addressed in Paper II.

\subsection{The interior metal distribution}
During Phase II the inner atmosphere is highly polluted, characterized by a high, super-solar metallicity. The extent of this region becomes more apparent when we plot profiles as function of mass instead of radius. In \fg{profiles} the metallicity and temperature profiles are plotted for several runs conducted at 0.2 au at three points during their evolution where the total mass reaches 5, 10, and 20 $M_\oplus$. In these figures the classical core, identified by $Z=1$, starts at $m=0$ while the remainder of the mass is dominated by the isothermal region 2b and the iso-metal region 2a, before a sharp drop-off in the metallicity in region 1. The temperature profile of the ``classical'' core (where $Z=1$, indicated light-red) reflects the values at the base of the envelope when the material became incorporated in Phase I. But as in our model heat is not being transported or generated in the core, this temperature profile does not affect the surrounding envelope. 

In \fg{profiles} the extent and value of the average metallicity over the regions where $Z>0.05$ -- our definition of the dilute core -- is indicated by the horizontal dashed line.  It is clear that with this definition the entire region 2 has become an extension of the core. We can expect that hotter envelopes are less dense, more polluted, and have therefore higher $Z_\mathrm{dc}$ values. In \fg{profiles} we observe this trend at fixed total mass. For example, in the middle column panels ($M=5\,M_\oplus$) higher $\dot{M}_P$ and $\kappa$ amounts to higher temperatures and higher $Z_\mathrm{dc}$.  Naturally, the dilute core metallicity $Z_\mathrm{dc}$ decreases with time as the convective region (2a) progresses outward and its metallicity decreases, reaching $\approx$0.5 at crossover for the $\dot{M}_\mathrm{peb}=10^{-5}\,M_\oplus\,\mathrm{yr}^{-1}$ runs. At crossover, $Z_\mathrm{dc}=0.5$ is the lowest value that can be attained, because it implies that the dilute core encompasses the entire envelope.
In \Tb{results} we have listed the value of $Z_\mathrm{dc}$ for all runs. As can be seen from \Tb{results} runs where significant vapor layers have built up, reach $Z_\mathrm{dc}\approx0.5$ at crossover, indicating that pebble accretion naturally produces a dilute core with approximately a 1:1 ratio between heavy metals and hydrogen/helium gas.

\section{Discussion}
\label{sec:discuss}
\subsection{The dilute core}
Recent analyses of Jupiter's gravitational moments as measured by the JUNO spacecraft indicate that Jupiter's central region is best characterized by a modest metallicity, which gradually transitions into the hydrogen/helium-dominated envelope -- the dilute core -- rather than a sharp interface between a fully metal core and a near metal-free envelope \citep{WahlEtal2017,DebrasChabrier2019}. Sublimation of solids during assembly naturally reproduces this trend and the region where the heavy metal fraction exceeds 5\% is substantial. Still, this fraction may not reach the levels as inferred from JUNO, which seem to indicate a dilute core metallicity as low $Z_\mathrm{dc}\approx0.3$ \citep{WahlEtal2017,LiuEtal2019}. \citet{LiuEtal2019} explain Jupiter's dilute core by means of an (head-on) impact of Jupiter's core with an embryo. Pebble accretion may offer an alternative or additional explanation. In this work, simulations end at crossover; we have not accounted for the effects of additional planetesimal accretion \citep{VenturiniHelled2020} nor for the runaway gas accretion and circumplanetary disk accretion phases, which may contribute to further solid accretion \citep{SzulagyiEtal2018,CilibrasiEtal2018,ShibataIkoma2019,ShibaikeEtal2019,PodolakEtal2020}, let alone model the interior profiles for billions of years. Redistribution of vapor by mixing, \eg\ due to a Ledoux-unstable configuration or miscibility of the core in metallic hydrogen \citep{WilsonMilitzer2012,SoubiranEtal2017}, or settling will affect $Z_\mathrm{dc}$ and the gravitational moments \citep{MuellerEtal2020}.  

In addition, planets that have not reached crossover -- (mini-)Neptunes -- will also be characterized by extended vapor regions after their formation. Although we did not consider ices, a gradual composition can also explain the measured properties of the solar system's ice-giants \citep{MarleyEtal1995,PodolakEtal2000,HelledEtal2011,VazanHelled2020}.  At any rate, our works provide the (realistic) profiles that emerge from the planet formation process by pebble accretion until crossover (\fg{profiles});  interior structure calculations can use these in their post-formation calculations to assess the long-term stability and evolution.

\subsection{Preservation of (sub)-Neptune planets}
One of the motivations of this study is to explain the presence of hydrogen/helium rich, but heavy metal dominant close-in planets -- sub-Neptunes -- from a pebble accretion perspective. Because of the inferred hydrogen/helium envelopes, it is natural to assume that these planets were formed within the gaseous disk. The challenge, then, is to understand why these planets stopped at the $\sim$sub-Neptune size and did not become hot-Jupiters. Delaying contraction by accretion of remnant planetesimals, as recently proposed for Jupiter \citep{AlibertEtal2018,GuileraEtal2020}, is not applicable at $\sim$0.1 au as these bodies are accreted on very short timescales. Another line of thinking is that these planets simply ran out of gas to accrete, because they formed late \citep[\eg][]{LeeChiang2016,OgiharaEtal2020} or the gas disks dispersed by photoevaporation, or because of inefficient accretion due to gap opening \citep{TanigawaIkoma2007,GinzburgChiang2019,VenturiniEtal2020ii}.

Most of the proposed solutions assume 
that atmospheres are dusty, which naturally prolongs the Kelvin-Helmholtz contraction. Here, we have adopted the opposite limit of entirely grain-free atmospheres. Our Phase III simulations and previous analytical work then indicate that the critical metal mass is reached already at $\approx$5 $M_\oplus$ (at 0.2 au) and much less further out.

Pebbles affect gas accretion in two ways. On the one hand, they accelerate gas accretion through accumulation of a vapor-rich atmosphere, which brings down the crossover mass by a couple of Earth masses. Similar to what \citet{VallettaHelled2020} concluded for \hho polluters, intake of nebular gas becomes significant before the pebble isolation mass is reached. Conversely,  by virtue of their small size, pebbles naturally provide an opacity, which reduces the intake of nebular gas. This pebble opacity is most relevant for the outer disk, however, as there the molecular opacity is low. When pebbles are small, $<$1\,mm, their contribution to $\kappa$ significantly raises the crossover mass (see \fg{pebble-opacity}). At $\sim$0.2 au, the opacity from pebbles could in principle delay gas accretion, but the window is rather narrow. The pebbles must be small, stay small (avoid coagulation) and the supply of pebbles must be uninterrupted. Otherwise, as we have seen in \fg{phase3}, atmospheres will quickly contract. 
Because at $\sim$0.1 au disk flows are strong both in terms of density and velocity, entropy advection or entire disk-atmosphere recycling appear to us the more natural mechanism through which mini-Neptunes can retain their atmospheres \citep{OrmelEtal2015i,FungEtal2015,CimermanEtal2017,LambrechtsLega2017,KurokawaTanigawa2018,BethuneRafikov2019i,MoldenhauerEtal2021}. 

\subsection{Importance of dust collision model}
Assigning a pebble opacity in the form of \eq{kappa-peb} is attractive because it can be directly linked to the pebble accretion rate and eliminates the need for an otherwise uncertain grain reduction factor.  However, \eq{kappa-peb} is only applicable when it is assumed that pebbles preserve their sizes. A motivation for this simple choice is that bouncing, among silicates, seems to be primarily a size-dependent phenomenon, as deduced from laboratory experiments among silicate particles \citep{WeidlingEtal2009,GuettlerEtal2010,ZsomEtal2010}.  Nevertheless, as the opacity is perhaps the single-most important factor in determining the evolutionary outcome, it is warranted to pursue a more sophisticated collisional process. Fragmentation and erosive collisions, either in the nebula \citep{ChenEtal2020} or in the envelope \citep{Ali-DibThompson2020,JohansenNordlund2020}, may contribute to a (dominating) dust opacity, but this trend will be strongly mediated by coagulation and sedimentation \citep{DullemondDominik2005,Mordasini2014,Ormel2014}. In an accompanying work we provide a more general opacity model by accounting for a fine-grain dust component, dust-pebble sweepup, and coagulation. \citep{BrouwersEtal2020}.

\subsection{Comparison to related works}
We briefly compare our work with \citet{BodenheimerEtal2018}, which also accounted for the absorption of metal vapor in the envelope as of planetesimal accretion. A side-by-side comparison cannot be done as \citet{BodenheimerEtal2018} study optimized their model towards understanding the properties of Kepler-36 c (at present), while we conduct a controlled parameter exploration to arrive at more general trends and conclusions.

Still, \citet{BodenheimerEtal2018} also end up with a core mass between 1-2 $M_\oplus$, consistent with our results. However, in their case, the subsequent buildup of a vapor layer (what we refer to as Phase II) is seen to suppress accretion of nebular gas, whereas in our model it is accelerated (see \fg{evol1}). We attribute this difference to the fact that in their work the vapor-rich layer is shielded from the remainder of the envelope. In their setup, planetesimals release their vapor in this layer (ablation being negligible). Therefore, the high-Z vapor layer essentially acts as a low density extension of the core, with the ``inflated'' core suppresses intake of nebular gas. In contrast, in our work the vapor is deposited much higher in the envelope and it naturally mixes with the rest of the envelope (in region 2a) as long as the overall luminosity stays positive. The higher molecular weight compresses these layers (the pressure gradient decreases), accelerating the intake of nebular gas as was shown by \citet{HoriIkoma2011} and \citet{VenturiniEtal2016} in case of \hho. But whereas these works assume a globally constant enrichment throughout the envelope, deposition of silicate vapor in our case occurs much deeper in the envelope (see \fg{Zsketch}).

\subsection{Model caveats and improvements}
One of the key features of our evolution model is the onset of an inner compositionally-frozen layer, which we model as isothermal. The motivation for this layer is physical; it stems from the fact that an insufficient amount of energy is available to lift the metal-rich lower layer with the metal-poor regions above. However, our choice to model this layer as isothermal is partly driven by the limitations of our numerical method. To proceed towards a more realistic treatment, a local model for energy transport (luminosity) as well as material transport (composition) is necessary. 
One approach for this is to opt for a pure local treatment for material and heat transport, based on the Ledoux convection criterion and mixing length theory. However, the Ledoux criterion only sets a lower bound for heat transport; when it indicates stability, whereas the Schwarzschield criterion does not, the region is dynamically stable but vibrationally unstable. Formally, a double-diffusion convection (DDC) model should be invoked \citep{MirouhEtal2012,LeconteChabrier2012,WoodEtal2013}. Including DDC in the models requires knowledge of thermodynamical properties which are poorly known (Prandtl number, diffusivity ratio, number of layers) for forming planets, which may affect the resulting interior structure.

In addition, interaction between core and inner envelope must be considered. In our model we have followed the classical choice that core and envelope are decoupled both thermally and materially. However, at these high densities and temperatures conduction may become relevant \citep{UmemotoEtal2006,StamenkovicEtal2011} and hydrogen from the envelope can be sequestered in the core \citep{KiteEtal2019}.

Another major limitation is the choice of a single species: \sioo. Even though a realistic EoS was employed, it lacked dissociation \citep{Melosh2007}. Inclusion of other volatile and metal species and their chemistry is necessary. In the outer disk, pebbles are expected to contain significant fractions of \hho ice, which will sublimate at much lower temperatures than \sioo. While the current consensus is of a 50-50\% refractory-to-volatile split for pebbles beyond the iceline \citep{Lodders2003,MorbidelliEtal2015}, the precise composition of the material that first accreted onto --say-- Jupiter is unknown. In addition, the rock-to-ice fraction may well vary among disks \citep{BitschBattistini2020}. Finally, when Jupiter would have formed near the iceline and ice sublimates readily in the outermost layer of the envelope, the \hho vapor may simply be advected back through the recycling mechanism \citep{OrmelEtal2015i,FungEtal2015}. Our assumption of a \hho vapor-free model for the 5 au runs, while far from ideal, is not without merits.

Conversely, more refractory rock species as MgO sublimate at higher temperatures \citep{Haynes2014}. Conceivably, it may result in an onion-like structure where volatiles readily sublimate and reach Phase II (sub-saturation), whereas the metal species are still accreting onto the core. 
Within the framework of the model, it is straightforward to add additional species, by following additional saturation vapor curves and using the ideal-EoS. However, non ideal effects, as well as chemical interactions between the different $Z$ species, are expected to affect the calculation and therefore require a proper description. These effects are only beginning to be treated by modelers \citep[\eg][]{HoriIkoma2011,HaldemannEtal2020} and we hope to address these topics in a future work.

\section{Conclusions}
\label{sec:conclude}
We have extended our results of Paper I, employing a real equation of state and an updated liquid-vapor curve.  The results for the first phase of direct core growth are similar, although our new calculations produce larger cores. The numerical calculations now proceed to the next phase, where the accreting pebbles fully evaporate, until the point where the crossover mass is reached.

Our main findings are the following:
\begin{enumerate}
    \item The critical metal mass -- the mass where the total mass in metals (core and envelope vapor) equals that of the hydrogen/helium gas -- heralds the onset of runaway gas accretion. Other findings based on (semi-)analytical arguments and on the ideal EoS, have also been numerically verified. The $M_\mathrm{core}$--$M_\mathrm{crossover}$ relation closely follows the expression derived in Paper II.
    \item The typical core mass that we find is around 1.5--2\,$M_\oplus$. This is slightly higher than found in Papers I and II and is primarily a consequence of the wet-adiabatic gradient we employed as well as a different liquid-vapor curve. 
    \item Temperature and metallicity \textit{profiles} differ significantly between the ideal and non-ideal EoS runs. As the long-term, post formation interior evolution of planets is sensitive to them, planet formation scenarios (\eg\ by pebble accretion) can be linked to the present state.
    \item The complete vaporization of pebbles naturally enriches the atmosphere in heavy elements. Based on elementary energy considerations, the dense, heavy metal layer may only partially mix with the H/He gas. Consequentially, a compositional gradient naturally forms. Typically, the dilute core metallicity approaches $Z_\mathrm{dc}\approx0.5$ upon reaching crossover.
    \item In the outer disk regions, where the molecular opacity is low, small pebbles dominate the opacity. When also the pebble accretion rate is high, atmospheres become very polluted. The crossover mass becomes large and its value independent of distance.
    \item When planets stop accreting pebbles (Phase III), accretion of nebular gas will quickly take over as the atmospheres loose their opacity. We therefore consider entropy advection or full-scale atmosphere recycling a more viable mechanism to prevent sub-Neptunes from reaching runaway gas accretion.
\end{enumerate}

\begin{acknowledgements}
    We thank the referee for a helpful report. We emphatically thank all participating members of the International Space Science Institute (ISSI) team ``Ice giants: formation, evolution and link to exoplanets'' for fruitful discussions. This work has benefited from the \texttt{matplotlib} \citep{Hunter2007}, \texttt{numpy} \citep{Numpy2020}, and \texttt{scipy} \citep{Scipy2020} open source software packages.
\end{acknowledgements}

\bibliography{phaseII}
\bibliographystyle{aa}

\clearpage
\newpage

\onecolumn

\addtocounter{table}{-1}
\begin{longtab}
    \centering
    \small
    \begin{longtable}{llllllllllllll}
        \caption{\label{tab:results}Conditions at crossover when the crossover mass $M_\mathrm{cross}$ is twice the mass in hydrogen and helium in the envelope $M_\mathrm{xy}$ and also (by definition) twice the mass in heavy metals in the core $M_\mathrm{core}$ and envelope $M_\mathrm{Z,env}$. The first five columns indicate the parameters: the distance from the star sets the boundary conditions (see \Tb{pars}) the pebble accretion rate, the equation of state (ideal else real), the pebble opacity given in terms of the pebble size (else, no pebble opacity). Additional parameters in the simpars column denote \textit{rec.}: simulations including entropy advection, where the outer boundary is reduced by 50\% (see text); III\textit{X}: phase III simulation, in which the total mass in solids is limited to \textit{X} $M_\oplus$; $\infty$: an infinite saturation pressure, ensuring all solids make it to the core and there is vapor in the envelope; \textit{n}: including radionuclide heating; \textit{S}: using the Stull liquid-vapor curve instead of the Kraus.  The table further list: the time of cross over $t_\mathrm{cross}$, the hydrogen/helium growth time $t_\mathrm{xy}$ at this point, the radius $r_{0.05}$ where the metal fraction has dropped to 5\%, and the average metallicity $Z_\mathrm{dc}$ of the said dilute core.} \\
    \hline\hline
$r$ & $\dot{M}_\mathrm{Peb}$ & EoS & simpars & $s_\mathrm{peb}$ & $M_\mathrm{core}$ & $M_\mathrm{cross}$ & $M_\mathrm{Z,env}$ & $M_\mathrm{xy}$ & $t_\mathrm{cross}$ & $t_\mathrm{xy}$ & $r_\mathrm{0.05}$ & $Z_\mathrm{dc}$ & Figure ref. \\
au & $M_\oplus\,\mathrm{yr}^{-1}$ &  &  & cm & $M_\oplus$ & $M_\oplus$ & $M_\oplus$ & $M_\oplus$ & Myr & Myr & $R_\oplus$ &  &  \\
\hline
\endfirsthead
\caption{continued.}\\
\hline\hline
$r$ & $\dot{M}_\mathrm{Peb}$ & EoS & simpars & $s_\mathrm{peb}$ & $M_\mathrm{core}$ & $M_\mathrm{cross}$ & $M_\mathrm{Z,env}$ & $M_\mathrm{xy}$ & $t_\mathrm{cross}$ & $t_\mathrm{xy}$ & $r_\mathrm{0.05}$ & $Z_\mathrm{dc}$ & Figure ref. \\
au & $M_\oplus\,\mathrm{yr}^{-1}$ &  &  & cm & $M_\oplus$ & $M_\oplus$ & $M_\oplus$ & $M_\oplus$ & Myr & Myr & $R_\oplus$ &  &  \\
\hline
\endhead
\hline
\endfoot
0.2  & {$10^{-4}$} &       &           &       & {$1.85$}     & {$20.7$}     & {$8.48$}     & {$10.3$}     & {$0.10$}     & {$0.067$}    & {$30.4$}     & {$0.51$}     &  \ref{fig:corecross}, \ref{fig:pebble-opacity} \\
0.2  & {$10^{-4}$} &       &           & 0.1   & {$1.72$}     & {$34.3$}     & {$15.4$}     & {$17.2$}     & {$0.17$}     & {$0.11$}     & {$54.0$}     & {$0.50$}     &  \ref{fig:pebble-opacity} \\
0.2  & {$10^{-5}$} & ideal &           &       & {$1.56$}     & {$13.1$}     & {$5.02$}     & {$6.51$}     & {$0.66$}     & {$0.68$}     & {$19.0$}     & {$0.52$}     &  \ref{fig:corecross} \\
0.2  & {$10^{-5}$} & ideal & rec.      &       & {$1.67$}     & {$16.5$}     & {$6.60$}     & {$8.24$}     & {$0.83$}     & {$1.13$}     & {$19.4$}     & {$0.51$}     &      \\
0.2  & {$10^{-5}$} &       &           &       & {$2.03$}     & {$15.2$}     & {$5.53$}     & {$7.63$}     & {$0.76$}     & {$0.36$}     & {$20.1$}     & {$0.52$}     &  \ref{fig:corecross}, \ref{fig:evol1}, \ref{fig:benchmark}, \ref{fig:pebble-opacity}, \ref{fig:profiles} \\
0.2  & {$10^{-5}$} &       & rec.      &       & {$2.12$}     & {$19.6$}     & {$7.71$}     & {$9.81$}     & {$0.98$}     & {$0.72$}     & {$22.1$}     & {$0.51$}     &  \ref{fig:benchmark}, \ref{fig:phase3} \\
0.2  & {$10^{-5}$} &       & III8      &       & {$2.03$}     & {$14.4$}     & {$5.10$}     & {$7.23$}     & {$0.81$}     & {$0.48$}     & {$18.4$}     & {$0.52$}     &  \ref{fig:phase3} \\
0.2  & {$10^{-5}$} &       & III6      &       & {$2.03$}     & {$11.9$}     & {$3.91$}     & {$5.96$}     & {$1.02$}     & {$1.23$}     & {$14.1$}     & {$0.54$}     &  \ref{fig:phase3} \\
0.2  & {$10^{-5}$} &       & III4      &       & {$2.04$}     & {$7.83$}     & {$1.92$}     & {$3.87$}     & {$5.28$}     & {$11.4$}     & {$7.62$}     & {$0.60$}     &  \ref{fig:phase3} \\
0.2  & {$10^{-5}$} & ideal & $\infty$  &       & {$32.3$}     & {$64.6$}     & {$0.0$}      & {$32.3$}     & {$3.23$}     & {$0.56$}     & {$3.82$}     & {$1.00$}     &      \\
0.2  & {$10^{-5}$} &       & $\infty$  &       & {$14.6$}     & {$28.8$}     & {$0.0$}      & {$14.2$}     & {$1.45$}     & {$0.061$}    & {$2.93$}     & {$1.00$}     &  \ref{fig:evol1}, \ref{fig:benchmark} \\
0.2  & {$10^{-5}$} &       &           & 1.0   & {$2.03$}     & {$15.1$}     & {$5.51$}     & {$7.53$}     & {$0.75$}     & {$0.36$}     & {$19.8$}     & {$0.52$}     &  \ref{fig:pebble-opacity} \\
0.2  & {$10^{-5}$} &       &           & 0.1   & {$1.75$}     & {$17.2$}     & {$6.82$}     & {$8.64$}     & {$0.86$}     & {$0.37$}     & {$22.3$}     & {$0.51$}     &  \ref{fig:pebble-opacity}, \ref{fig:phase3}, \ref{fig:profiles} \\
0.2  & {$10^{-5}$} &       & III8      & 0.1   & {$1.75$}     & {$15.0$}     & {$5.73$}     & {$7.51$}     & {$0.90$}     & {$0.49$}     & {$19.0$}     & {$0.52$}     &  \ref{fig:phase3} \\
0.2  & {$10^{-5}$} &       & III6      & 0.1   & {$1.75$}     & {$11.9$}     & {$4.19$}     & {$5.96$}     & {$1.11$}     & {$1.31$}     & {$13.8$}     & {$0.54$}     &  \ref{fig:phase3} \\
0.2  & {$10^{-5}$} &       & III4      & 0.1   & {$1.75$}     & {$7.90$}     & {$2.21$}     & {$3.95$}     & {$5.68$}     & {$11.4$}     & {$7.66$}     & {$0.60$}     &  \ref{fig:phase3} \\
0.2  & {$10^{-6}$} & ideal &           &       & {$1.77$}     & {$9.54$}     & {$3.03$}     & {$4.74$}     & {$4.80$}     & {$3.24$}     & {$12.4$}     & {$0.53$}     &  \ref{fig:evol0}, \ref{fig:analysis}, \ref{fig:corecross} \\
0.2  & {$10^{-6}$} & ideal & rec.      &       & {$1.96$}     & {$12.1$}     & {$4.10$}     & {$6.07$}     & {$6.06$}     & {$5.66$}     & {$12.9$}     & {$0.51$}     &      \\
0.2  & {$10^{-6}$} &       &           &       & {$2.20$}     & {$11.2$}     & {$3.38$}     & {$5.58$}     & {$5.58$}     & {$1.85$}     & {$13.0$}     & {$0.54$}     &  \ref{fig:standard}, \ref{fig:evol0}, \ref{fig:analysis}, \ref{fig:corecross}, \ref{fig:benchmark}, \ref{fig:pebble-opacity}, \ref{fig:profiles} \\
0.2  & {$10^{-6}$} &       & rec.      &       & {$2.38$}     & {$14.7$}     & {$4.99$}     & {$7.36$}     & {$7.37$}     & {$3.94$}     & {$14.6$}     & {$0.51$}     &  \ref{fig:benchmark} \\
0.2  & {$10^{-6}$} &       & III8      &       & {$2.20$}     & {$11.1$}     & {$3.31$}     & {$5.61$}     & {$5.71$}     & {$1.96$}     & {$12.8$}     & {$0.54$}     &      \\
0.2  & {$10^{-6}$} &       & III6      &       & {$2.20$}     & {$10.6$}     & {$3.09$}     & {$5.33$}     & {$5.97$}     & {$2.58$}     & {$12.0$}     & {$0.55$}     &      \\
0.2  & {$10^{-6}$} &       & III4      &       & {$2.21$}     & {$8.03$}     & {$1.76$}     & {$4.06$}     & {$9.84$}     & {$10.9$}     & {$7.75$}     & {$0.59$}     &      \\
0.2  & {$10^{-6}$} &       & n         &       & {$2.19$}     & {$11.0$}     & {$3.33$}     & {$5.44$}     & {$5.52$}     & {$1.96$}     & {$12.9$}     & {$0.54$}     &      \\
0.2  & {$10^{-6}$} & ideal & S         &       & {$1.34$}     & {$7.90$}     & {$2.64$}     & {$3.92$}     & {$3.98$}     & {$3.34$}     & {$13.0$}     & {$0.52$}     &  \ref{fig:corecross} \\
0.2  & {$10^{-6}$} &       & S         &       & {$1.50$}     & {$9.16$}     & {$3.11$}     & {$4.55$}     & {$4.61$}     & {$2.42$}     & {$14.6$}     & {$0.53$}     &  \ref{fig:corecross} \\
0.2  & {$10^{-6}$} & ideal & $\infty$  &       & {$14.5$}     & {$28.8$}     & {$0.0$}      & {$14.3$}     & {$14.5$}     & {$2.28$}     & {$2.92$}     & {$1.00$}     &      \\
0.2  & {$10^{-6}$} &       & $\infty$  &       & {$8.84$}     & {$17.6$}     & {$0.0$}      & {$8.80$}     & {$8.83$}     & {$0.54$}     & {$2.48$}     & {$1.00$}     &  \ref{fig:benchmark} \\
0.2  & {$10^{-6}$} &       &           & 1.0   & {$2.20$}     & {$11.2$}     & {$3.38$}     & {$5.57$}     & {$5.58$}     & {$1.84$}     & {$13.0$}     & {$0.54$}     &  \ref{fig:pebble-opacity} \\
0.2  & {$10^{-6}$} &       &           & 0.1   & {$2.22$}     & {$11.2$}     & {$3.40$}     & {$5.56$}     & {$5.62$}     & {$1.86$}     & {$13.1$}     & {$0.54$}     &  \ref{fig:pebble-opacity} \\
0.2  & {$10^{-6}$} &       &           & 0.01  & {$1.72$}     & {$20.5$}     & {$8.43$}     & {$10.3$}     & {$10.1$}     & {$3.95$}     & {$22.2$}     & {$0.50$}     &  \ref{fig:pebble-opacity} \\
0.2  & {$10^{-7}$} & ideal &           &       & {$1.92$}     & {$7.21$}     & {$1.69$}     & {$3.60$}     & {$36.1$}     & {$13.8$}     & {$8.44$}     & {$0.55$}     &  \ref{fig:corecross} \\
0.2  & {$10^{-7}$} & ideal & rec.      &       & {$2.21$}     & {$9.26$}     & {$2.40$}     & {$4.66$}     & {$46.1$}     & {$25.7$}     & {$8.98$}     & {$0.51$}     &      \\
0.2  & {$10^{-7}$} &       &           &       & {$2.39$}     & {$8.41$}     & {$1.81$}     & {$4.21$}     & {$42.0$}     & {$8.17$}     & {$8.69$}     & {$0.57$}     &  \ref{fig:corecross}, \ref{fig:benchmark}, \ref{fig:pebble-opacity} \\
0.2  & {$10^{-7}$} &       & rec.      &       & {$2.67$}     & {$11.0$}     & {$2.90$}     & {$5.48$}     & {$55.7$}     & {$20.6$}     & {$9.76$}     & {$0.53$}     &  \ref{fig:benchmark} \\
0.2  & {$10^{-7}$} & ideal & $\infty$  &       & {$7.20$}     & {$14.3$}     & {$0.0$}      & {$7.15$}     & {$71.7$}     & {$10.0$}     & {$2.31$}     & {$1.00$}     &      \\
0.2  & {$10^{-7}$} &       & $\infty$  &       & {$5.42$}     & {$10.9$}     & {$0.0$}      & {$5.45$}     & {$54.2$}     & {$4.19$}     & {$2.11$}     & {$1.00$}     &  \ref{fig:benchmark} \\
0.2  & {$10^{-7}$} &       &           & 1.0   & {$2.39$}     & {$8.41$}     & {$1.81$}     & {$4.21$}     & {$42.0$}     & {$8.19$}     & {$8.69$}     & {$0.57$}     &  \ref{fig:pebble-opacity} \\
0.2  & {$10^{-7}$} &       &           & 0.1   & {$2.39$}     & {$8.37$}     & {$1.81$}     & {$4.17$}     & {$42.0$}     & {$8.25$}     & {$8.65$}     & {$0.57$}     &  \ref{fig:pebble-opacity} \\
1    & {$10^{-4}$} &       &           &       & {$2.07$}     & {$13.6$}     & {$4.65$}     & {$6.84$}     & {$0.067$}    & {$0.021$}    & {$19.2$}     & {$0.55$}     &  \ref{fig:corecross}, \ref{fig:pebble-opacity} \\
1    & {$10^{-4}$} &       &           & 0.1   & {$1.78$}     & {$45.1$}     & {$20.9$}     & {$22.5$}     & {$0.23$}     & {$0.12$}     & {$77.4$}     & {$0.51$}     &  \ref{fig:pebble-opacity} \\
1    & {$10^{-5}$} & ideal &           &       & {$1.67$}     & {$8.98$}     & {$2.85$}     & {$4.46$}     & {$0.45$}     & {$0.25$}     & {$13.5$}     & {$0.56$}     &  \ref{fig:corecross} \\
1    & {$10^{-5}$} & ideal & rec.      &       & {$1.76$}     & {$11.7$}     & {$4.02$}     & {$5.88$}     & {$0.58$}     & {$0.40$}     & {$16.7$}     & {$0.52$}     &      \\
1    & {$10^{-5}$} &       &           &       & {$2.13$}     & {$10.2$}     & {$2.95$}     & {$5.11$}     & {$0.51$}     & {$0.12$}     & {$12.9$}     & {$0.57$}     &  \ref{fig:corecross}, \ref{fig:benchmark}, \ref{fig:pebble-opacity} \\
1    & {$10^{-5}$} &       & rec.      &       & {$2.24$}     & {$12.6$}     & {$4.10$}     & {$6.22$}     & {$0.63$}     & {$0.21$}     & {$15.9$}     & {$0.54$}     &  \ref{fig:benchmark}, \ref{fig:phase3} \\
1    & {$10^{-5}$} &       & III8      &       & {$2.13$}     & {$10.0$}     & {$2.89$}     & {$4.97$}     & {$0.51$}     & {$0.14$}     & {$12.5$}     & {$0.58$}     &  \ref{fig:phase3} \\
1    & {$10^{-5}$} &       & III6      &       & {$2.13$}     & {$9.77$}     & {$2.78$}     & {$4.86$}     & {$0.53$}     & {$0.18$}     & {$12.1$}     & {$0.58$}     &  \ref{fig:phase3} \\
1    & {$10^{-5}$} &       & III4      &       & {$2.13$}     & {$8.01$}     & {$1.83$}     & {$4.04$}     & {$0.70$}     & {$0.47$}     & {$8.89$}     & {$0.62$}     &  \ref{fig:phase3} \\
1    & {$10^{-5}$} & ideal & $\infty$  &       & {$11.0$}     & {$22.2$}     & {$0.0$}      & {$11.2$}     & {$1.10$}     & {$0.074$}    & {$2.67$}     & {$1.00$}     &      \\
1    & {$10^{-5}$} &       & $\infty$  &       & {$7.66$}     & {$15.5$}     & {$0.0$}      & {$7.80$}     & {$0.77$}     & {$0.027$}    & {$2.36$}     & {$1.00$}     &  \ref{fig:benchmark} \\
1    & {$10^{-5}$} &       &           & 1.0   & {$2.14$}     & {$10.2$}     & {$2.98$}     & {$5.12$}     & {$0.51$}     & {$0.14$}     & {$13.0$}     & {$0.57$}     &  \ref{fig:pebble-opacity} \\
1    & {$10^{-5}$} &       &           & 0.1   & {$1.80$}     & {$17.9$}     & {$7.16$}     & {$8.92$}     & {$0.90$}     & {$0.27$}     & {$23.7$}     & {$0.52$}     &  \ref{fig:pebble-opacity}, \ref{fig:phase3} \\
1    & {$10^{-5}$} &       & III8      & 0.1   & {$1.80$}     & {$14.9$}     & {$5.58$}     & {$7.50$}     & {$0.87$}     & {$0.22$}     & {$19.5$}     & {$0.54$}     &  \ref{fig:phase3} \\
1    & {$10^{-5}$} &       & III6      & 0.1   & {$1.81$}     & {$11.9$}     & {$4.09$}     & {$5.97$}     & {$0.83$}     & {$0.20$}     & {$14.9$}     & {$0.57$}     &  \ref{fig:phase3} \\
1    & {$10^{-5}$} &       & III4      & 0.1   & {$1.81$}     & {$7.93$}     & {$2.16$}     & {$3.96$}     & {$0.89$}     & {$0.55$}     & {$8.55$}     & {$0.65$}     &  \ref{fig:phase3} \\
1    & {$10^{-6}$} & ideal &           &       & {$1.73$}     & {$6.83$}     & {$1.70$}     & {$3.39$}     & {$3.44$}     & {$1.15$}     & {$9.16$}     & {$0.58$}     &  \ref{fig:corecross} \\
1    & {$10^{-6}$} & ideal & rec.      &       & {$1.87$}     & {$8.79$}     & {$2.53$}     & {$4.39$}     & {$4.40$}     & {$2.13$}     & {$11.5$}     & {$0.54$}     &      \\
1    & {$10^{-6}$} &       &           &       & {$2.21$}     & {$7.69$}     & {$1.66$}     & {$3.82$}     & {$3.87$}     & {$0.82$}     & {$8.72$}     & {$0.61$}     &  \ref{fig:corecross}, \ref{fig:benchmark}, \ref{fig:pebble-opacity} \\
1    & {$10^{-6}$} &       & rec.      &       & {$2.34$}     & {$9.98$}     & {$2.60$}     & {$5.05$}     & {$4.93$}     & {$1.04$}     & {$11.3$}     & {$0.55$}     &  \ref{fig:benchmark} \\
1    & {$10^{-6}$} & ideal & $\infty$  &       & {$6.51$}     & {$13.1$}     & {$0.0$}      & {$6.60$}     & {$6.51$}     & {$0.39$}     & {$2.24$}     & {$1.00$}     &      \\
1    & {$10^{-6}$} &       & $\infty$  &       & {$4.97$}     & {$9.87$}     & {$0.0$}      & {$4.90$}     & {$4.97$}     & {$0.24$}     & {$2.05$}     & {$1.00$}     &  \ref{fig:benchmark} \\
1    & {$10^{-6}$} &       &           & 1.0   & {$2.21$}     & {$7.72$}     & {$1.65$}     & {$3.85$}     & {$3.87$}     & {$0.60$}     & {$8.72$}     & {$0.61$}     &  \ref{fig:pebble-opacity} \\
1    & {$10^{-6}$} &       &           & 0.1   & {$2.26$}     & {$8.67$}     & {$2.05$}     & {$4.37$}     & {$4.31$}     & {$0.64$}     & {$10.2$}     & {$0.59$}     &  \ref{fig:pebble-opacity} \\
1    & {$10^{-6}$} &       &           & 0.01  & {$1.78$}     & {$21.6$}     & {$8.93$}     & {$10.8$}     & {$10.7$}     & {$3.52$}     & {$25.7$}     & {$0.51$}     &  \ref{fig:pebble-opacity} \\
1    & {$10^{-7}$} & ideal &           &       & {$1.73$}     & {$5.20$}     & {$0.89$}     & {$2.58$}     & {$26.2$}     & {$5.54$}     & {$6.26$}     & {$0.62$}     &  \ref{fig:corecross} \\
1    & {$10^{-7}$} & ideal & rec.      &       & {$1.93$}     & {$6.77$}     & {$1.48$}     & {$3.36$}     & {$34.1$}     & {$10.9$}     & {$8.02$}     & {$0.56$}     &      \\
1    & {$10^{-7}$} &       &           &       & {$2.25$}     & {$5.84$}     & {$0.63$}     & {$2.95$}     & {$28.9$}     & {$2.31$}     & {$5.55$}     & {$0.69$}     &  \ref{fig:corecross}, \ref{fig:benchmark}, \ref{fig:pebble-opacity} \\
1    & {$10^{-7}$} &       & rec.      &       & {$2.45$}     & {$7.78$}     & {$1.43$}     & {$3.90$}     & {$38.8$}     & {$4.79$}     & {$7.84$}     & {$0.59$}     &  \ref{fig:benchmark} \\
1    & {$10^{-7}$} & ideal & $\infty$  &       & {$3.90$}     & {$7.72$}     & {$0.0$}      & {$3.83$}     & {$38.9$}     & {$2.96$}     & {$1.89$}     & {$1.00$}     &      \\
1    & {$10^{-7}$} &       & $\infty$  &       & {$3.25$}     & {$6.54$}     & {$0.0$}      & {$3.29$}     & {$32.5$}     & {$1.82$}     & {$1.78$}     & {$1.00$}     &  \ref{fig:benchmark} \\
1    & {$10^{-7}$} &       &           & 1.0   & {$2.25$}     & {$5.85$}     & {$0.64$}     & {$2.96$}     & {$28.9$}     & {$2.32$}     & {$5.57$}     & {$0.68$}     &  \ref{fig:pebble-opacity} \\
1    & {$10^{-7}$} &       &           & 0.1   & {$2.25$}     & {$5.76$}     & {$0.64$}     & {$2.86$}     & {$29.0$}     & {$2.67$}     & {$5.54$}     & {$0.69$}     &  \ref{fig:pebble-opacity} \\
5    & {$10^{-4}$} &       &           &       & {$2.03$}     & {$8.91$}     & {$2.45$}     & {$4.43$}     & {$0.045$}    & {$0.011$}    & {$11.3$}     & {$0.63$}     &  \ref{fig:corecross}, \ref{fig:pebble-opacity} \\
5    & {$10^{-4}$} &       &           & 0.1   & {$1.80$}     & {$47.6$}     & {$21.9$}     & {$23.9$}     & {$0.24$}     & {$0.10$}     & {$81.1$}     & {$0.52$}     &  \ref{fig:pebble-opacity} \\
5    & {$10^{-5}$} & ideal &           &       & {$1.60$}     & {$5.86$}     & {$1.31$}     & {$2.95$}     & {$0.29$}     & {$0.078$}    & {$7.98$}     & {$0.65$}     &  \ref{fig:corecross} \\
5    & {$10^{-5}$} & ideal & rec.      &       & {$1.68$}     & {$7.44$}     & {$2.03$}     & {$3.73$}     & {$0.37$}     & {$0.13$}     & {$10.6$}     & {$0.58$}     &      \\
5    & {$10^{-5}$} &       &           &       & {$2.07$}     & {$6.44$}     & {$1.17$}     & {$3.20$}     & {$0.32$}     & {$0.035$}    & {$7.00$}     & {$0.70$}     &  \ref{fig:corecross}, \ref{fig:benchmark}, \ref{fig:pebble-opacity} \\
5    & {$10^{-5}$} &       & rec.      &       & {$2.14$}     & {$8.27$}     & {$2.00$}     & {$4.13$}     & {$0.41$}     & {$0.088$}    & {$9.84$}     & {$0.61$}     &  \ref{fig:benchmark}, \ref{fig:phase3} \\
5    & {$10^{-5}$} &       & III8      &       & {$2.07$}     & {$6.40$}     & {$1.16$}     & {$3.17$}     & {$0.32$}     & {$0.030$}    & {$6.96$}     & {$0.70$}     &  \ref{fig:phase3} \\
5    & {$10^{-5}$} &       & III6      &       & {$2.07$}     & {$6.44$}     & {$1.15$}     & {$3.23$}     & {$0.33$}     & {$0.027$}    & {$6.93$}     & {$0.70$}     &  \ref{fig:phase3} \\
5    & {$10^{-5}$} &       & III4      &       & {$2.07$}     & {$6.23$}     & {$1.05$}     & {$3.11$}     & {$0.33$}     & {$0.047$}    & {$6.62$}     & {$0.71$}     &  \ref{fig:phase3} \\
5    & {$10^{-5}$} & ideal & $\infty$  &       & {$4.97$}     & {$10.0$}     & {$0.0$}      & {$5.04$}     & {$0.50$}     & {$0.024$}    & {$2.05$}     & {$1.00$}     &      \\
5    & {$10^{-5}$} &       & $\infty$  &       & {$4.07$}     & {$8.04$}     & {$0.0$}      & {$3.98$}     & {$0.41$}     & {$0.023$}    & {$1.91$}     & {$1.00$}     &  \ref{fig:benchmark} \\
5    & {$10^{-5}$} &       &           & 1.0   & {$2.09$}     & {$7.25$}     & {$1.57$}     & {$3.59$}     & {$0.37$}     & {$0.044$}    & {$8.32$}     & {$0.67$}     &  \ref{fig:pebble-opacity} \\
5    & {$10^{-5}$} &       &           & 0.1   & {$1.91$}     & {$18.1$}     & {$7.07$}     & {$9.09$}     & {$0.90$}     & {$0.31$}     & {$23.9$}     & {$0.52$}     &  \ref{fig:pebble-opacity}, \ref{fig:phase3} \\
5    & {$10^{-5}$} &       & III8      & 0.1   & {$1.91$}     & {$14.7$}     & {$5.43$}     & {$7.31$}     & {$0.86$}     & {$0.21$}     & {$19.5$}     & {$0.54$}     &  \ref{fig:phase3} \\
5    & {$10^{-5}$} &       & III6      & 0.1   & {$1.92$}     & {$11.7$}     & {$3.96$}     & {$5.80$}     & {$0.80$}     & {$0.15$}     & {$14.9$}     & {$0.59$}     &  \ref{fig:phase3} \\
5    & {$10^{-5}$} &       & III4      & 0.1   & {$1.93$}     & {$7.95$}     & {$2.04$}     & {$3.98$}     & {$0.64$}     & {$0.066$}    & {$8.80$}     & {$0.70$}     &  \ref{fig:phase3} \\
5    & {$10^{-6}$} & ideal &           &       & {$1.55$}     & {$4.07$}     & {$0.51$}     & {$2.01$}     & {$2.06$}     & {$0.17$}     & {$4.75$}     & {$0.74$}     &  \ref{fig:corecross} \\
5    & {$10^{-6}$} & ideal & rec.      &       & {$1.68$}     & {$5.49$}     & {$1.08$}     & {$2.74$}     & {$2.76$}     & {$0.63$}     & {$7.08$}     & {$0.63$}     &      \\
5    & {$10^{-6}$} &       &           &       & {$2.01$}     & {$4.63$}     & {$0.30$}     & {$2.32$}     & {$2.31$}     & {$0.071$}    & {$4.17$}     & {$0.81$}     &  \ref{fig:corecross}, \ref{fig:benchmark}, \ref{fig:pebble-opacity} \\
5    & {$10^{-6}$} &       & rec.      &       & {$2.17$}     & {$6.09$}     & {$0.86$}     & {$3.07$}     & {$3.03$}     & {$0.31$}     & {$6.20$}     & {$0.68$}     &  \ref{fig:benchmark} \\
5    & {$10^{-6}$} & ideal & $\infty$  &       & {$2.87$}     & {$5.75$}     & {$0.0$}      & {$2.87$}     & {$2.87$}     & {$0.18$}     & {$1.70$}     & {$1.00$}     &      \\
5    & {$10^{-6}$} &       & $\infty$  &       & {$2.49$}     & {$4.95$}     & {$0.0$}      & {$2.47$}     & {$2.49$}     & {$0.15$}     & {$1.62$}     & {$1.00$}     &  \ref{fig:benchmark} \\
5    & {$10^{-6}$} &       &           & 1.0   & {$2.01$}     & {$4.69$}     & {$0.32$}     & {$2.36$}     & {$2.33$}     & {$0.14$}     & {$4.22$}     & {$0.80$}     &  \ref{fig:pebble-opacity} \\
5    & {$10^{-6}$} &       &           & 0.1   & {$2.22$}     & {$7.59$}     & {$1.57$}     & {$3.80$}     & {$3.79$}     & {$0.38$}     & {$8.59$}     & {$0.64$}     &  \ref{fig:pebble-opacity} \\
5    & {$10^{-6}$} &       &           & 0.01  & {$1.89$}     & {$21.1$}     & {$8.76$}     & {$10.5$}     & {$10.6$}     & {$3.38$}     & {$25.4$}     & {$0.52$}     &  \ref{fig:pebble-opacity} \\
5    & {$10^{-7}$} & ideal &           &       & {$1.35$}     & {$3.00$}     & {$0.15$}     & {$1.50$}     & {$15.0$}     & {$0.85$}     & {$3.08$}     & {$0.83$}     &  \ref{fig:corecross} \\
5    & {$10^{-7}$} & ideal & rec.      &       & {$1.59$}     & {$4.00$}     & {$0.41$}     & {$2.00$}     & {$20.0$}     & {$2.41$}     & {$4.51$}     & {$0.70$}     &      \\
5    & {$10^{-7}$} &       &           &       & {$1.52$}     & {$3.07$}     & {$0.021$}    & {$1.53$}     & {$15.4$}     & {$1.01$}     & {$2.14$}     & {$0.94$}     &  \ref{fig:corecross}, \ref{fig:benchmark}, \ref{fig:pebble-opacity} \\
5    & {$10^{-7}$} &       & rec.      &       & {$2.02$}     & {$4.42$}     & {$0.18$}     & {$2.22$}     & {$22.0$}     & {$0.84$}     & {$3.70$}     & {$0.80$}     &  \ref{fig:benchmark} \\
5    & {$10^{-7}$} & ideal & $\infty$  &       & {$1.68$}     & {$3.34$}     & {$0.0$}      & {$1.66$}     & {$16.8$}     & {$1.29$}     & {$1.43$}     & {$1.00$}     &      \\
5    & {$10^{-7}$} &       & $\infty$  &       & {$1.55$}     & {$3.08$}     & {$0.0$}      & {$1.54$}     & {$15.4$}     & {$1.04$}     & {$1.39$}     & {$1.00$}     &  \ref{fig:benchmark} \\
5    & {$10^{-7}$} &       &           & 1.0   & {$1.52$}     & {$3.07$}     & {$0.021$}    & {$1.53$}     & {$15.4$}     & {$1.01$}     & {$2.14$}     & {$0.94$}     &  \ref{fig:pebble-opacity} \\
5    & {$10^{-7}$} &       &           & 0.1   & {$1.64$}     & {$3.32$}     & {$0.040$}    & {$1.64$}     & {$16.8$}     & {$1.01$}     & {$2.46$}     & {$0.91$}     &  \ref{fig:pebble-opacity} \\
\hline
\end{longtable}
\end{longtab}

\end{document}